\documentclass[iop]{emulateapj}
\usepackage{amsmath,amstext}
\usepackage{hyperref}
\usepackage[all]{hypcap}
\usepackage{color}

\newif\ifemulateapj
\makeatletter%
\@ifclassloaded{emulateapj}%
{\emulateapjtrue}%
{\emulateapjfalse}
\makeatother%

\newcommand{\kmsmpc}{\ensuremath{\mathrm{\ km\ s^{-1}\ Mpc^{-1}}}}

\shorttitle{CSP-I: Absolute Calibration and the Hubble Constant}
\shortauthors{Burns et al. (2018)}
\begin{document}
\title{
The Carnegie Supernova Project: Absolute Calibration and the Hubble Constant
}

\ifemulateapj
   \author{
Christopher~R.~Burns\altaffilmark{1},
Emilie~Parent\altaffilmark{2},
M.~M.~Phillips\altaffilmark{3},
Maximilian~Stritzinger\altaffilmark{4},
Kevin~Krisciunas\altaffilmark{5},
Nicholas~B.~Suntzeff\altaffilmark{5},
Eric~Y.~Hsiao\altaffilmark{6},
Carlos~Contreras\altaffilmark{3},
Jorge~Anais\altaffilmark{3},
Luis~Boldt\altaffilmark{3},
Luis~Busta\altaffilmark{3},
Abdo~Campillay\altaffilmark{3},
Sergio~Castell\'{o}n\altaffilmark{3},
Gast\'{o}n~Folatelli\altaffilmark{3,7},
Wendy~L.~Freedman\altaffilmark{1},
Consuelo~Gonz\'{a}lez\altaffilmark{3},
Mario~Hamuy\altaffilmark{8},
Peter~Heoflich\altaffilmark{6},
Wojtek~Krzeminski\altaffilmark{3,9},
Barry~F.~Madore\altaffilmark{1},
Nidia~Morrell\altaffilmark{3},
S.~E.~Persson\altaffilmark{1},
Miguel~Roth\altaffilmark{3},
Francisco~Salgado\altaffilmark{3},
Jacqueline~Ser\'{o}n\altaffilmark{3,10} and 
Sim\'{o}n~Torres\altaffilmark{3,11}
}
\altaffiltext{1}{Observatories of the Carnegie Institution for Science, 813 Santa Barbara St, Pasadena, CA, 91101, USA}
\altaffiltext{2}{Department~of Physics and McGill Space Institute, McGill University, Montreal, QC H3A 2T8, Canada}
\altaffiltext{3}{Carnegie Institution of Washington, Las Campanas Observatory, Casilla 601, Chile}
\altaffiltext{4}{Department of Physics and Astronomy, Aarhus University, Ny Munkegade 120, DK-8000 Aarhus C, Denmark}
\altaffiltext{5}{George P. and Cynthia Woods Mitchell Institute for Fundamental Physics and Astronomy, Texas A\&M University, Department of Physics and Astronomy, College Station, TX, 77843, USA}
\altaffiltext{6}{Department of Physics, Florida State University, Tallahassee, FL 32306, USA}
\altaffiltext{7}{Facultad de Ciencias Astron\'{o}micas y Geof\'{i}sicas, Universidad Nacional de La Plata, Instituto de Astrof\'{i}sica de La Plata (IALP), CONICET, Paseo del Bosque S/N, B1900FWA La Plata, Argentina}
\altaffiltext{8}{Universidad de Chile, Departmento de Astronomia, Casilla 36-D, Santiago, Chile}
\altaffiltext{9}{Deceased}
\altaffiltext{10}{Cerro Tololo Inter-American Observatory, Casilla 603, La Serena, Chile}
\altaffiltext{11}{SOAR Telescope, Casilla 603, La Serena, Chile}

\else
   \input{authors_aastex6.tex}
\fi

\begin{abstract}
We present an analysis of the final data release of the 
\textit{Carnegie Supernova Project I}, focusing on  the absolute calibration 
of the luminosity-decline-rate
relation for Type Ia supernovae (SNe~Ia) using new intrinsic color relations 
with respect to the
color-stretch parameter, $s_{BV}$, enabling improved dust extinction
corrections. We investigate to what degree the so-called fast-declining SNe~Ia 
   can be used to determine accurate extragalactic distances. 
We 
estimate 
the intrinsic scatter
in the luminosity-decline-rate relation, and find it ranges from
$\pm 0.13$ mag to $\pm 0.18$ mag with
   no obvious dependence on wavelength.
Using the Cepheid variable star data from the SH0ES project
\citep{Riess:2016}, the SN~Ia distance scale  is calibrated and the 
Hubble constant is estimated using our optical \textit{and} 
near-infrared sample, 
and these results are compared to those determined exclusively from a 
near-infrared sub-sample. 
The systematic
effect of the supernova's host galaxy mass is investigated as a function 
of wavelength and is found to decrease toward redder wavelengths, suggesting
this effect may be due to dust properties of the host. Using estimates of
the dust extinction derived from optical and NIR wavelengths, and applying
these to $H$ band, we derive a Hubble constant $H_0 = 73.2 +/- 2.3 \kmsmpc$,
whereas using a simple $B-V$ color-correction applied to $B$ band yields
$H_0 = 72.7 +/- 2.1 \kmsmpc$.
Photometry of two calibrating
SNe~Ia from the CSP-II sample, SN~2012ht and SN~2015F, is presented and
used to improve the calibration of the SN~Ia distance ladder.
\end{abstract}

\keywords{supernovae: general, cosmology: cosmological parameters, ISM:dust, 
extinction}

\section{Introduction}

\label{sec:Intro} The successful use of Type~Ia supernovae (SNe~Ia) as
extragalactic distance indicators hinges on the discovery that the rate of
evolution of their light-curves (i.e., decline rate) is correlated with their 
intrinsic luminosity
\citep{Pskovskii:1977,Phillips:1993}.  Since this initial discovery a handful
of different parameters have been used to characterize the decline rate,
including $\Delta m_{15}(B)$ \citep{Phillips:1993}, the light-curve stretch
\citep{Perlmutter:1999}, the Multi-Color Light Curve Shapes (MLCS) parameter
$\Delta$ \citep{Riess:1996,Jha:2007}, and the Spectral Adaptive Light curve
Template (SALT) stretch-like parameter $x_1$ \citep{Guy:2005}. One problem
common to all these parameters, however, is their difficulty in working with
what seems to be a separate class of SNe~Ia, the so-called fast-decliners, which
are often conflated with the spectrally classified 1991bg-like objects
\citep{Filippenko:1992,Leibundgut:1993}. In this paper, we use the recently
proposed color-stretch parameter, $s_{BV}$, \citep{Burns:2014} as a way forward
to deal with fast-declining SNe~Ia.

\begin{figure*}
   \plotone{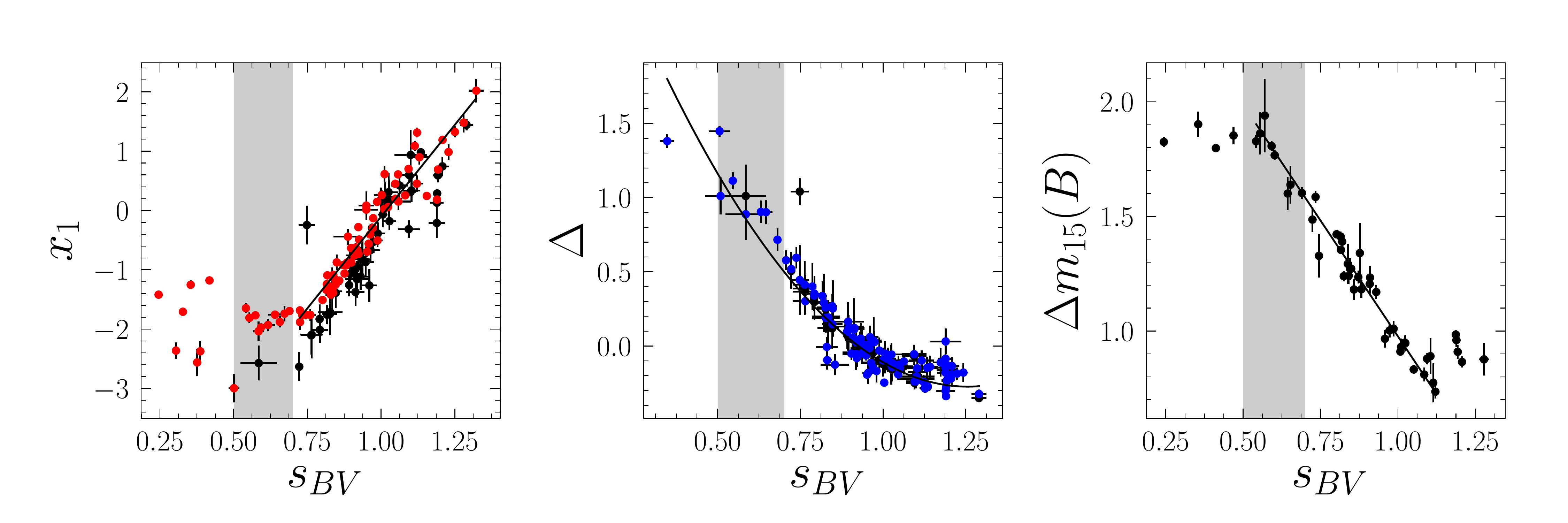}
   \caption{
      \label{fig:shape_parameters}
      Comparison of different light-curve parameters as a function of
      the color stretch $s_{BV}$. A low value of $s_{BV}$ indicates a
      faster-declining, and hence less luminous SN. (Left) The SALT2 $x_1$
      parameter. Black points are from \citet{Hicken:2009}, 
      red points are from \citet{Burns:2014}. (Middle) The
      MLCS2k2 $\Delta$ parameter.  Black points are from
      \citet{Hicken:2009}, blue points are from
      \citet{Jha:2007}. (Right) The decline rate $\Delta m_{15}(B)$ from
      the CSP-I DR3 sample. In all three panels, the relations from
      equations \ref{eq:x1-sbv} - \ref{eq:dm15-sbv} are shown as 
      solid black lines. Grey rectangles show the range of
      $s_{BV}$ where the simple relations with respect to 
      $x_1$ and $\Delta m_{15}(B)$ break down and one might consider 
      to be the transition region between normal and fast-declining SNe~Ia.}
\end{figure*}

Another key ingredient to using SNe~Ia to infer distances is the ability to
correct for dust extinction. The simplest approach is to perform a
one-parameter color correction, as reddening is directly proportional to
extinction \citep{Cardelli:1989,Fitzpatrick:1999}. This was first done by
\citet{Tripp:1998} and we refer to it hereafter as the Tripp method.  The
disadvantage of this approach is that it allows one to determine only a
combination of luminosity and a color (a reddening-free luminosity). While this
is treated as a nuisance parameter in cosmology, the luminosities of SNe~Ia are 
important to
determine in order to study the physics of their explosion mechanism(s)
\citep{Ashall:2014,Hoeflich:2017}.  In a
previous paper \citep{Burns:2014}, we developed a method to determine
more accurate extinction corrections.
In this paper we apply this method to calibrate the luminosity-decline-rate
relation \citep{Phillips:1999} and compare the results with the calibration
based on the Tripp method.

One of the major goals of the \textit{Carnegie Supernova Project I} (hereafter
CSP-I; \citealt{Hamuy:2006}) was to produce a photometrically homogeneous set
of SNe~Ia of exceptional quality. 
While an obvious application of such a sample
is for cosmology and measuring the Hubble constant $H_0$, 
the high cadence and small photometric errors of CSP-I was not necessary for
a statistical sample to anchor the Hubble diagram.
Indeed, the publication history of the CSP is
skewed much more toward the physics of the SNe and their environments, rather
than constraining cosmological parameters. Now that the final data release
(DR3) of the CSP-I is published \citep[see][]{Krisciunas:2017}, we present  an
analysis of the intrinsic luminosities of the entire sample and a
derivation of $H_0$ tied to a Cepheid distance scale. This paper is a
continuation of previous work on the intrinsic colors of SNe~Ia which allowed
us to properly deal with dust extinction and produce an improved
luminosity-decline-rate relation. As part of that analysis we use  the
color-stretch parameter $s_{BV}$ \citep{Burns:2014}.  As we demonstrate below,
$s_{BV}$ is a more reliable
light-curve parameter
when fitting the fast-declining SNe~Ia
(i.e., those for which $\Delta m_{15}(B) > 1.7$ mag), both in 
predicting the shapes of the optical and near-infrared (NIR) light curves, and
in producing better behaved intrinsic colors as a function of decline rate. We
shall now forge ahead in using $s_{BV}$  as an intrinsic SN~Ia luminosity
indicator.

In the preliminary analysis paper \citep{Folatelli:2010}, we used an 
assumed value of the Hubble constant in order to calibrate the luminosities
of the CSP-I sample. In this paper, we use Cepheid variables from the 
SH0ES project \citep{Riess:2016} to solve for the absolute calibration of
the SN~Ia luminosities as a function of decline-rate and measure the Hubble
constant. In particular, we investigate the systematics involved in using
different reddening corrections, different sub-samples of SNe~Ia, and
different wavelength ranges (e.g., optical versus NIR).

The paper is organized follows. Section 2 briefly describes the CSP-I sample
and the photometric system used. Section 3 reviews $s_{BV}$  and compares it
with other existing parameters. Section 4 presents the absolute calibration of
the CSP-I DR3 for fixed values of the Hubble constant, $H_0$, using both 
a simple color-correction and a method that estimates a proper dust extinction. 
In section 5, Cepheid data 
from the SH0ES project \citep{Riess:2016} are used to calibrate
the distances to 19 SN~Ia host galaxies and to determine
$H_0$.   Section~6 concludes with a summary of our results.

\section{The CSP-I DR3 sample} 
\label{sec:DR3_sample} 
The final CSP-I data release is  presented by \citet{Krisciunas:2017}.  For the
purposes of this paper, we briefly describe the sample, underlining important
aspects for this work.

DR3 consists of 134 SNe~Ia 
observed between 2004 and 2009. The majority of the objects were
observed in optical ($ugriBV$) and NIR ($YJH$) passbands, and  at least some
visual-wavelength spectra were obtained for most of the objects
\citep[e.g.,][]{Folatelli:2010}.  The sample contains objects with redshifts
spanning the range of $0.004 < z < 0.083$. The CSP-I was a purely follow-up
program of SNe discovered by other surveys using the facilities of Las 
Campanas Observatory with the goal of
minimizing systematics due to calibration and extinction. Our primary source
of objects was the Lick Observatory Supernova Survey \citep{Filippenko:2001}
and, being a targeted survey, is biased toward luminous hosts.

The  philosophy for the CSP-I observations was ``quality over quantity''. We
chose to follow-up fewer objects, but with higher cadence and signal-to-noise
than most other follow-up projects. This  allowed us to construct accurate
light-curve template models as a function of decline rate \citep{Burns:2011},
while the wide wavelength coverage allowed us to model the intrinsic
colors and extinctions \citep{Burns:2014}. The CSP-I also produced
high-fidelity filter functions using a monochrometer \citep{Rheault:2010,
Stritzinger:2011}, which improves S- and K-corrections \citep{Stritzinger:2005}
and allows accurate determination of absolute zero-points for the 
photometric natural system  \citep{Krisciunas:2017}. The net result was a 
high-quality set of SNe~Ia 
classified by decline rate and having redshift- and
extinction-corrected light curves. These are crucial for determining distances
and producing accurate Hubble diagrams.

\subsection{Sample Used for Cosmology}
Of the 134 objects in the CSP-I DR3 sample, 123 are {\em bona-fide} SNe~Ia. 
The remaining 11 objects are members of peculiar sub-types and have been 
omitted from our analysis:
\begin{itemize}
   \item SN~2005hk, SN~2008ae, SN~2008ha, SN2009J and SN~2010ae
         are all 2002cx-like SNe \citep{Li:2001};
   \item SN~2007if and SN~2009dc are Super-Chandrasekhar (SC) candidates
         \citep{Howell:2006};
         
   \item SN~2006bt and SN~2006ot are peculiar objects and form their own
         sub-group \citep{Foley:2010, Stritzinger:2011} and are relatively 
         easy to identify with NIR photometry \citep{Phillips:2012};
   \item SN~2005gj and SN~2008J are 2002ic-like SNe~Ia \citep{Hamuy:2003}, that 
   exhibit signatures of interaction (i.e., prevalent Balmer emission lines)
   produced  from the SN ejecta shocking circumstellar material. 
\end{itemize}

We also eliminate three normal SNe~Ia:  SN~2006dd, SN~2007so and
SN~2008bd, whose CSP-I observations begin well after maximum, when their light
curves are in their linear decline phase, and for which template light-curve
fits are unreliable.  In summary, this leaves us with 120 CSP-I SNe~Ia to  
use in our analysis. In order to anchor the SN~Ia distance ladder, we also
use 14 SNe~Ia from the literature.
Finally, 3 SNe~Ia from the CSP-II project 
\citep{Phillips:2018} have Cepheid distances and are included in this
analysis. The photometry for SN~2012fr is published in 
\citet{Contreras:2018}, while the photometry for SN~2012ht and SN~2015F are 
presented in appendix \ref{sec:12ht+15F}. This brings the total number
of SN~Ia anchors to 19, and is the same set used by \citet{Riess:2016}.

Each supernova is fit with light-curve templates using the SNooPy
\citep{Burns:2011} package, yielding estimates of $s_{BV}$ and the 
maximum brightness $m_X$ in each filter $X$. SNooPy also applies
K-corrections to remove the effects of red-shift. The pseudo-colors
derived from the maximum brightnesses are used to estimate extinctions
using the methods of \citet{Burns:2014} (see section \ref{sec:color_reddening}).
Visual representations of the fits of all the CSP-I SNe are presented
in \citet{Krisciunas:2017}.

\begin{figure*}
   \plotone{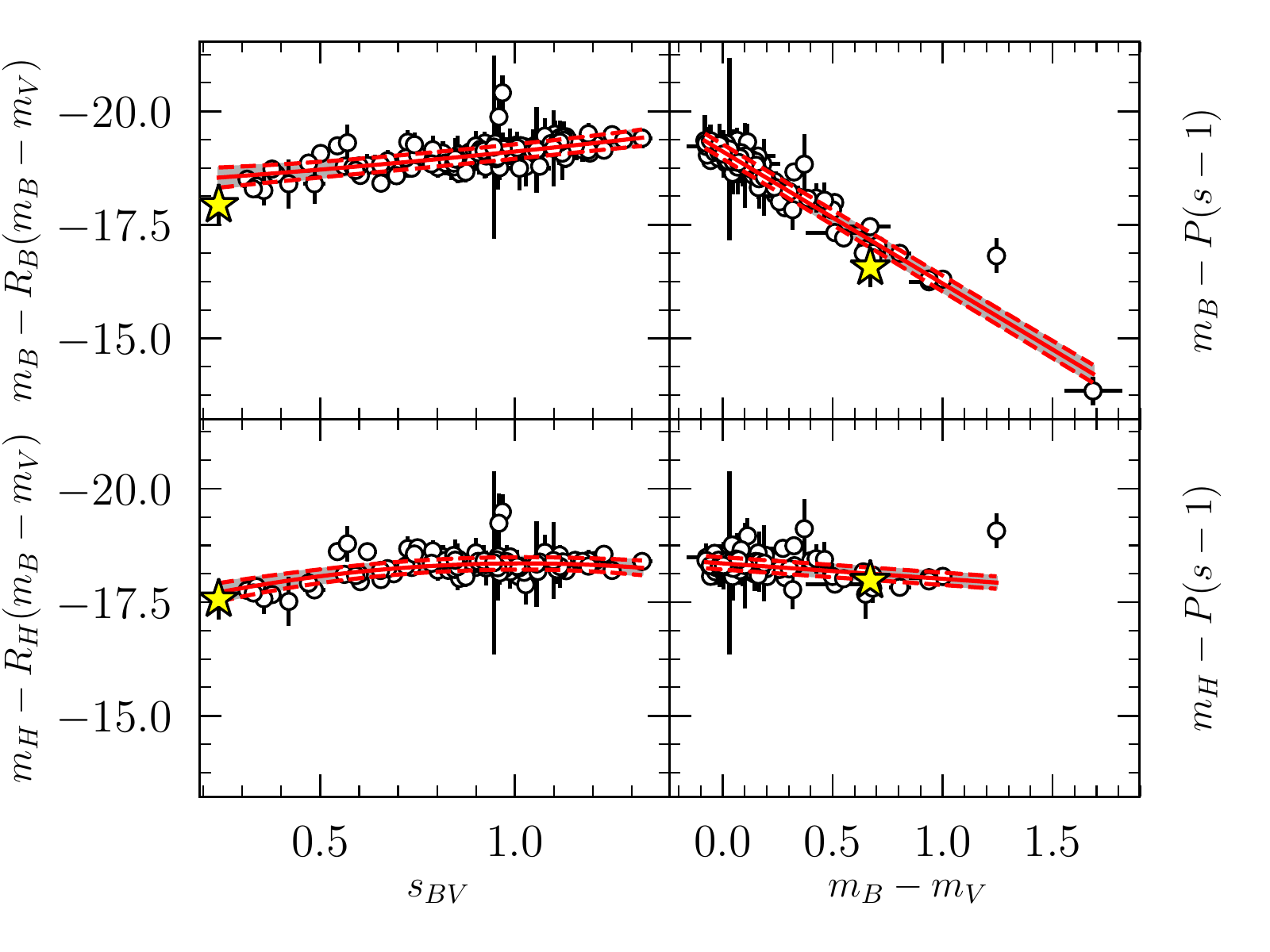}
   \caption{
   \label{fig:Tripp_BBV_HBV}
   Comparison of the Tripp calibration in $B$ and $H$ correcting by $(B-V)$
   color.  The top row shows the $B$ band and the bottom row shows the $H$
   band.  The left column shows absolute magnitudes corrected for color as a
   function of $s_{BV}$, while the right column shows absolute magnitudes
   corrected for $s_{BV}$  as a function of $(B-V)$ color. The red solid lines
   show the best-fit polynomial and the dashed red lines show the $\pm
   \sigma_X$ intrinsic dispersion. 
   The yellow star corresponds to
   SN~2006mr, which is the fastest declining object in our sample
   and whose distance is inferred from three normal SNe~Ia
   that were hosted by the same galaxy, NGC~1316.}
\end{figure*}

\section{The Color-Stretch Parameter $s_{BV}$}

The use of $s_{BV}$ was motivated by the problem $\Delta m_{15}(B)$ has in
measuring the decline rate of the fastest declining objects.  Once beyond
$\Delta m_{15}(B) \sim 1.7$ mag, the transition from the initial decline to the
linear decline occurs before day 15, which causes $\Delta m_{15}(B)$ to be less
sensitive to the rate of evolution of the $B$-band light curve
\citep{Phillips:2012,Burns:2014}. This transition was initially used by
\citet{Pskovskii:1977} as a way to define a decline rate for SNe~Ia, however it
proved too impractical and was never adopted by others. \citet{Hoflich:2010} 
also attempted to fix the problem by applying a stretch to the light-curves,
and then measuring a ''stretched`` $\Delta m_{15,s}$.

Another very pronounced feature is the time at which the SN~Ia reaches its
reddest color.
For the well-behaved $B-V$ colors, this usually takes place $\sim 30$ days 
after $B$ maximum, but
occurs earlier for fast-decliners and later
for slow-decliners. Together with
the time of maximum light for the SN~Ia, this provides a kind of clock for
measuring how fast the object is evolving. For convenience, we use the time
between $B$ maximum and $(B-V)$ maximum, but other filters could also be used.
Dividing by 30 days yields a stretch-like parameter for which $s_{BV}\sim 1$
for ``normal" SNe~Ia while fast decliners typically have $s_{BV} < 0.5$. Unlike
$\Delta m_{15}(B)$, $s_{BV}$ is insensitive to extinction 
\citep{Phillips:1999}. And unlike $\Delta m_{15}(B)$, the
correlation between time of $(B-V)$ maximum and the shape of the optical and
NIR light curves does not break down for the fast decliners \citep{Burns:2014}.
This is a significant improvement over the templates introduced in the first
version of SNooPy \citep{Burns:2011}, which used $\Delta m_{15}(B)$ as a
light-curve shape parameter.

There are theoretical reasons to believe that $s_{BV}$ might prove to be a
better diagnostic of the intrinsic brightness of SNe~Ia. The location of the
peak $(B-V)$ color is generally thought to be due to the recombination of
\ion{Fe}{3} to \ion{Fe}{2}, which deposits energy into the ejecta, making it
bluer \citep{Kasen:2006, Hoeflich:2017} 
and producing the secondary maxima
in the NIR bands.
Recombination occurs when the ejecta
have cooled to a particular temperature and the temporal phase when this occurs
depends on the total energy deposited into the ejecta and the time-dependent
opacity, both of which depend on the amount of $^{56}$Ni generated during the
thermonuclear disruption \citep[see, e.g.,][]{Hoeflich:2017}.  The amount of
$^{56}$Ni itself is also thought to be the primary determinant of the
luminosity of the SN~Ia \citep{Arnett:1982}.

\subsection{Comparison with Other Light Curve Parameters}

We briefly consider how $s_{BV}$ compares with the other most commonly-used
light curve parameters: $\Delta m_{15}(B)$, $\Delta$, and $x_1$. In particular
we are interested in analytic formulas that enable the conversion between one
light-curve parameter to another, but also a measure of the RMS
(root-mean-square) scatter in these relations. While template light-curve
fitters can measure statistical errors using standard methods when fitting for
the decline rate, they tend to be very small due to the precision of the 
photometry
and the large number of points being fit by a single-parameter function.
Determining the systematic error is not so obvious. However, if we compare
several different estimators of the decline rate, the scatter will give us an
indication of the systematic error introduced by the fitting process.

Comparing $s_{BV}$ with $\Delta m_{15}(B)$ is straightforward, as these are
direct measurements from the $B$ and $V$ light curves themselves. We use the
light-curve  analysis package SNooPy \citep{Burns:2011} to interpolate each
light curve with a Gaussian Process interpolator
\citep{Rasmussen:2006}. These are used to  measure 
the time of $B$ maximum, $\Delta m_{15}(B)$, and the epoch of $(B-V)$
maximum. We also use SNooPy to compute and apply  K-corrections
\citep{Oke:1968} for each light curve using the \citet{Hsiao:2007} spectral
template. In the right panel of Figure \ref{fig:shape_parameters} we plot the
results and a linear fit over the range $0.6 < s_{BV} < 1.2$. Interestingly, it
appears that $\Delta m_{15}(B)$ flattens out at the slow-declining end as
well as at the fast end. Wherever these relations flatten out (or go vertical)
is a regime where one parameter is potentially telling us more than the other
and could therefore prove to be a better discriminator of the decline rate.

For the other two commonly-used shape parameters, we use the values of $\Delta$
and $x_1$ published by their authors. For $x_1$ we also used SALT2 to fit some
CSP-I objects to show how $x_1$ fails in the same way as $\Delta m_{15}(B)$:
at high decline rates, the relation flattens out\footnote{To be fair, SALT2 was
never meant to be used to fit fast-decliners.}. In contrast, the MLCS2k2
parameter $\Delta$ shows a very clear correlation for the fast-decliners, but
begins to flatten out at the slow end. This is to be expected, since $\Delta$
is defined as a brightness correction relative to a ``standard" SN~Ia, so the
middle panel of Figure \ref{fig:shape_parameters} is simply a scaled version of
the luminosity-decline-rate relation in $V$ band. In no case do any of these
relations become vertical:  $s_{BV}$ seems to be more informative than the
other three parameters. Nevertheless, it is useful to be able to compare these
parameters, and so we derive analytic relations between 
$s_{BV}$ and each of the other three parameters. This
is done using simple $\chi^2$-minimization.  
Fitting a
linear relation for $x_1$, we obtain:
\begin{equation}
   \label{eq:x1-sbv}
   x_1 = -0.10(0.03) + 6.2(0.2)\left(s_{BV}-1\right).
\end{equation}
This relation is valid for $s_{BV} > 0.7$. The RMS dispersion is $0.27$ in 
$x_1$ or $0.04$ in $s_{BV}$. 

Unlike $x_1$,
the relation between $\Delta$ and $s_{BV}$ is quite continuous over the
entire range and shows no obvious break point. Nevertheless, the relation
flattens at larger $s_{BV}$ and so we fit a quadratic:
\begin{equation}
   \label{eq:delta-sbv}
   \Delta = -0.11(0.02) - 1.28(0.08)\left(s_{BV}-1\right) + 
      2.5(0.2)\left(s_{BV}-1\right)^2,
\end{equation}
which is is valid 
for the entire range of $s_{BV}$.
The RMS dispersion 
is 0.11 mag in $\Delta$ or 0.08 in $s_{BV}$. 

Finally, the relation between $s_{BV}$ and $\Delta m_{15}(B)$ is
found to be 
\begin{equation}
   \label{eq:dm15-sbv}
   \Delta m_{15}(B) = 0.98(0.01) - 2.02(0.05) \left(s_{BV}-1\right),
\end{equation}
and this is valid over the range $0.5 < s_{BV} < 1.15$. The RMS dispersion is 
$0.06$ magnitudes in $\Delta m_{15}(B)$ or 0.03 in $s_{BV}$. 

Given these three
independent measures of the decline rate for SNe~Ia,
we can take the average
RMS of the fits to equations \ref{eq:x1-sbv} - \ref{eq:dm15-sbv} 
as indicative of the systematic error in $s_{BV}$ for any {\em one} object, 
which
is $\sigma (s_{BV}) = 0.05$. 
When fitting the decline-rate relation for 
cosmological purposes, this will become a random error added in quadrature to 
the statistical errors reported by SNooPy.

\section{Intrinsic Luminosities}
\label{sec:Intrinsic-Luminosities}

Having presented  $s_{BV}$, which quantifies the relative
luminosity of SNe~Ia, we now turn to the other parameter needed to determine 
distances:  the extinction.
A commonly used technique to handle the extinction is the Wesenheit function 
\citep{Madore:1982}.
In \citet{Folatelli:2010}, we presented this calibration as the Tripp-method 
and will continue to
use this name. The advantage of this method is that for a fixed value of $R_V$,
the correction removes the effects of extinction without needing to know the
intrinsic colors. However, the assumption of constant $R_V$ 
(also labeled $\beta$ in
other analyses 
\citep[e.g.][]{Guy:2007})
is demonstrably not
correct in our own Milky-Way galaxy
\citep{Cardelli:1989,Fitzpatrick:1999,Nataf:2015}, as well as in the host
galaxies of SNe~Ia \citep{Riess:1996,Mandel:2011,Burns:2014}. Furthermore, 
if one is
interested in the intrinsic luminosity of SNe~Ia, then a proper treatment of
the extinction is necessary, including variations in the reddening curve from
host to host. In this section we discuss the inference of the reddening
correction, the luminosity-decline-rate relation, and the possibility of 
using the
fast-declining SNe~Ia as standardizable candles.

\subsection{Tripp Calibration}
\label{sec:Tripp_calibration}

For a single set of three filters labeled $X$,$Y$,$Z$, which define a magnitude
in band $X$ and $Y-Z$ color, the Tripp method models the observed
peak magnitude
$m_{X}$ as

\begin{eqnarray}
\label{eq:Tripp_model}
m_{X} & = & P_{XYZ}^{N}(s_{BV}-1)+\mu\left(z_{cmb}, H_0, C\right)+ \\
      & & R_{XYZ}\left(m_{Y}-m_{Z}\right) + 
          \alpha_M\left(\log_{10}M_*/M_\odot-M_0\right)
       \nonumber
\end{eqnarray}
where $P_{XYZ}^{N}(s_{BV}-1)$ is a polynomial of order $N$ as a function
of $s_{BV}-1$, and
$\mu = \mu\left(z_{cmb},H_0,C\right)$ is the distance modulus given a set of
cosmological parameters $C$ including $H_0$.  $R_{XYZ}$ can
be interpreted either as a simple parameter to be determined in the fitting, or
if one assumes a reddening law, 
\begin{equation}
   R_{XYZ} = \frac{R_X}{R_Y - R_Z},
   \label{eq:RXYZ}
\end{equation}
where each term is a function only of $R_V$ through the reddening law
\citep[e.g.][]{Fitzpatrick:1999,Cardelli:1989}. A special case, 
the combination
$XYZ = BBV$, yields $R_{BBV} = R_V$ owing to the fact that 
$R_B = R_V + 1$. Using equation \ref{eq:RXYZ} , one can fit multiple
filter triplet combinations simultaneously and solve for a single $R_V$.
The final term of equation \ref{eq:Tripp_model} takes into account the
correlation between the host galaxy stellar mass and intrinsic luminosity
of its SN~Ia \citep{Neill:2009,Sullivan:2010,Kelly:2010,Uddin:2017}
with $\alpha_M$ being the slope of the correlation and $M_0$ an arbitrary
mass zero-point, which we take to be $M_0 = 10^{11} M_\odot$. We derive
host stellar masses for the CSP-I sample in Appendix \ref{sec:HostMass}.
It is important to point out, though, that any estimate of host mass
will involve the distance to the host. This introduces a serious co-variance
in host mass with Hubble residual and must be handled carefully when
doing inference using equation \ref{eq:Tripp_model}. 
Specifically, since $\log_{10}(M_*/M_\odot) \propto 0.4\mu$ (see appendix
\ref{sec:HostMass}), the covariance will be
$cov(\mu,log_{10}(M_*/M_\odot)) = 0.4\delta \mu^2$, where $\delta \mu$ is
the error in distance.
We take this into account explicitly
by including the distance-dependence in $\log_{10}M_*/M_\odot$ (see
equation \ref{eq:stellar_mass}). 

A serious drawback of this approach is that both intrinsic (i.e. physics
of the SN explosion) and extrinsic (dust extinction) sources
of color variation are conflated into a single correction and so the
inferred value of $R_V$ cannot reflect the true average dust properties. 
Recent work by
\citet{Mandel:2017} shows that these effects can be separated in a 
statistical sense, alleviating the bias introduced in determining $R_V$.
For the purposes of this paper, we will not attempt to separate these
effects in the Tripp method, but rather tackle the problem by using more 
sophisticated color models to properly separate reddening and intrinsic
color variations (see section \ref{sec:color_reddening}).

% table:  Tripp Fit Parameters\label{tab:Tripp_coef}
\capstartfalse
\begin{deluxetable*}{lllllllll}
\tabletypesize{\footnotesize}
\tablewidth{0pc}
\tablecolumns{9}
\tablecaption{Tripp Fit Parameters\label{tab:Tripp_coef}}
\tablehead{
   \colhead{X} & \colhead{$P^0$} & 
   \colhead{$P^1$} & \colhead{$P^2$} & \colhead{$R_{XBV}$} & \colhead{$R_V$} & \colhead{$\alpha$} & \colhead{$\sigma_{X}$} & \colhead{$v_{pec}$} \\
 \colhead{} & \colhead{mag} & \colhead{mag} & \colhead{mag} & \colhead{} & \colhead{} & \colhead{mag/dex} & \colhead{mag} & \colhead{$\mathrm{km \cdot s^{-1}}$}  }
\startdata
\sidehead{Full sample}
$B$ & $-$19.182(062) & $-$0.89(11) & $-$0.02(30) & 2.81(09) & 1.65(08) & $-$0.063(031) & 0.13 & 310\\
$V$ & $-$19.181(061) & $-$0.89(11) & $-$0.02(30) & 1.82(09) & 1.65(08) & $-$0.063(030) & 0.13 & 310\\
$u$ & $-$18.818(097) & $-$1.28(17) & 0.32(44) & 3.64(13) & 1.13(52) & $-$0.135(051) & 0.22 & 233\\
$g$ & $-$19.229(084) & $-$0.90(11) & $-$0.13(31) & 2.38(09) & 1.57(09) & $-$0.073(032) & 0.13 & 317\\
$r$ & $-$19.099(059) & $-$0.74(10) & 0.38(27) & 1.38(08) & 1.78(08) & $-$0.077(028) & 0.12 & 302\\
$i$ & $-$18.523(059) & $-$0.48(10) & 0.41(27) & 0.98(08) & 1.85(09) & $-$0.081(028) & 0.12 & 295\\
$Y$ & $-$18.517(077) & $-$0.07(11) & 1.19(30) & 0.42(09) & 1.34(21) & $-$0.083(032) & 0.12 & 284\\
$J$ & $-$18.633(062) & $-$0.37(12) & 0.61(32) & 0.36(10) & 1.27(36) & $-$0.056(029) & 0.11 & 336\\
$H$ & $-$18.431(062) & $-$0.05(12) & 1.18(31) & 0.27(09) & 1.28(57) & $-$0.050(030) & 0.11 & 299\\
\sidehead{$B-V < 0.5$}
$B$ & $-$19.161(062) & $-$0.94(11) & $-$0.36(43) & 2.70(15) & 1.54(14) & $-$0.053(030) & 0.13 & 238\\
$V$ & $-$19.161(061) & $-$0.94(11) & $-$0.37(44) & 1.70(15) & 1.54(14) & $-$0.052(030) & 0.13 & 238\\
$u$ & $-$18.793(095) & $-$1.35(18) & $-$0.47(64) & 3.63(14) & 1.12(51) & $-$0.117(050) & 0.21 & 197\\
$g$ & $-$19.206(082) & $-$0.97(11) & $-$0.57(43) & 2.28(15) & 1.48(14) & $-$0.064(032) & 0.13 & 245\\
$r$ & $-$19.081(060) & $-$0.77(10) & 0.12(41) & 1.27(14) & 1.67(13) & $-$0.069(028) & 0.13 & 233\\
$i$ & $-$18.501(060) & $-$0.52(10) & $-$0.10(41) & 0.93(14) & 1.79(17) & $-$0.072(029) & 0.13 & 245\\
$Y$ & $-$18.497(076) & $-$0.10(11) & 0.34(41) & 0.57(15) & 1.69(35) & $-$0.070(031) & 0.12 & 222\\
$J$ & $-$18.601(062) & $-$0.43(11) & $-$0.42(45) & 0.43(16) & 1.51(58) & $-$0.047(029) & 0.11 & 284\\
$H$ & $-$18.400(062) & $-$0.10(12) & 0.17(47) & 0.27(14) & 1.33(85) & $-$0.046(030) & 0.11 & 248\\
\sidehead{$s_{BV} > 0.5$}
$B$ & $-$19.159(062) & $-$0.93(12) & $-$0.61(43) & 2.80(09) & 1.64(09) & $-$0.053(030) & 0.13 & 329\\
$V$ & $-$19.159(061) & $-$0.94(11) & $-$0.62(43) & 1.80(09) & 1.64(09) & $-$0.052(031) & 0.13 & 328\\
$u$ & $-$18.790(097) & $-$1.32(18) & $-$0.35(70) & 3.60(11) & 1.10(45) & $-$0.122(052) & 0.22 & 232\\
$g$ & $-$19.204(084) & $-$0.96(12) & $-$0.80(43) & 2.37(10) & 1.56(09) & $-$0.064(033) & 0.13 & 334\\
$r$ & $-$19.081(060) & $-$0.77(11) & $-$0.05(39) & 1.36(08) & 1.76(08) & $-$0.069(029) & 0.12 & 317\\
$i$ & $-$18.499(059) & $-$0.52(10) & $-$0.21(38) & 0.96(08) & 1.82(10) & $-$0.071(028) & 0.12 & 309\\
$Y$ & $-$18.480(076) & $-$0.11(11) & 0.32(42) & 0.35(09) & 1.18(22) & $-$0.076(031) & 0.11 & 280\\
$J$ & $-$18.593(060) & $-$0.44(12) & $-$0.35(45) & 0.29(10) & 1.02(36) & $-$0.048(029) & 0.11 & 330\\
$H$ & $-$18.394(061) & $-$0.10(12) & 0.13(47) & 0.19(08) & 0.82(52) & $-$0.046(030) & 0.11 & 295\\
\sidehead{$s_{BV} > 0.5$ and $B-V < 0.5$}
$B$ & $-$19.162(061) & $-$0.94(11) & $-$0.30(46) & 2.71(15) & 1.55(14) & $-$0.053(030) & 0.13 & 241\\
$V$ & $-$19.163(061) & $-$0.94(11) & $-$0.31(46) & 1.71(15) & 1.55(14) & $-$0.053(031) & 0.13 & 240\\
$u$ & $-$18.796(095) & $-$1.35(17) & $-$0.42(69) & 3.63(14) & 1.12(51) & $-$0.118(050) & 0.21 & 198\\
$g$ & $-$19.207(083) & $-$0.96(11) & $-$0.53(46) & 2.28(16) & 1.48(15) & $-$0.064(032) & 0.13 & 247\\
$r$ & $-$19.083(060) & $-$0.77(10) & 0.17(42) & 1.28(14) & 1.68(13) & $-$0.069(029) & 0.13 & 236\\
$i$ & $-$18.501(061) & $-$0.52(10) & $-$0.10(43) & 0.92(15) & 1.78(17) & $-$0.072(029) & 0.13 & 248\\
$Y$ & $-$18.489(075) & $-$0.10(10) & 0.15(42) & 0.53(15) & 1.59(35) & $-$0.073(031) & 0.12 & 217\\
$J$ & $-$18.598(063) & $-$0.43(12) & $-$0.48(47) & 0.41(16) & 1.48(57) & $-$0.047(029) & 0.11 & 284\\
$H$ & $-$18.395(061) & $-$0.11(12) & 0.03(48) & 0.26(14) & 1.24(84) & $-$0.046(030) & 0.11 & 246\\
\enddata
\tablecomments{$P^0$, $P^1$, and $P^2$ are the coefficients of the zeroth,
first, and second order terms of the polynomial $P^N_{XYZ}\left(s_{BV}-1\right)$
from equation \ref{eq:Tripp_model}.}
\end{deluxetable*}
\capstarttrue

\begin{figure*}
   \plotone{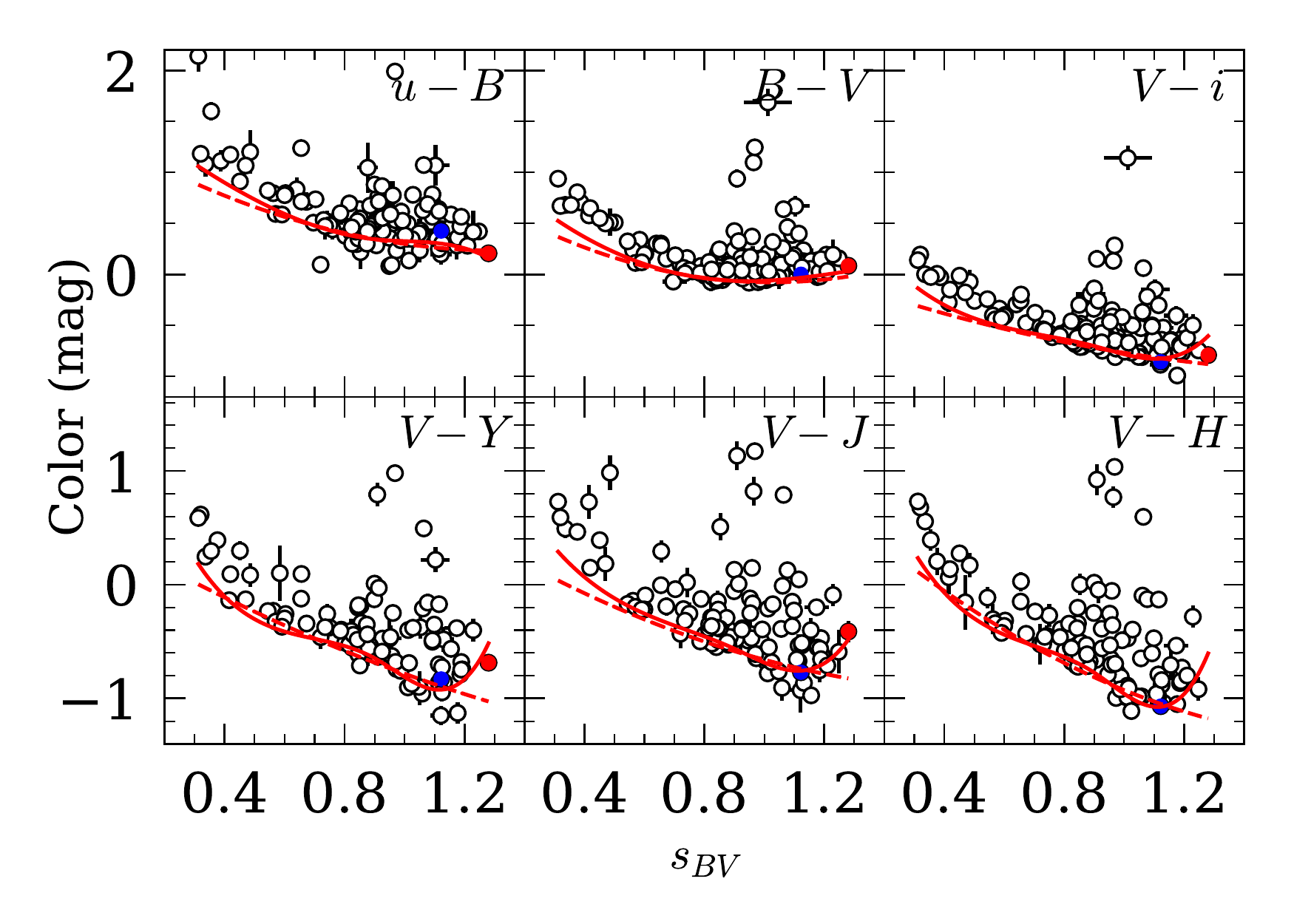}
   \caption{
   \label{fig:color_panels}
   Plot of observed colors of the CSP-I SNe~Ia DR3 sample. Extinction
   can only cause points to increase in color, therefore the intrinsic
   colors are defined by the blue edges of these distributions. The models
   for these intrinsic colors are plotted as red lines.
   The solid red line 
   represents the best-fit spline function of  $s_{BV}$, whereas
   the dashed red line is the polynomial fit from \citet{Burns:2014}.
   The red point corresponds to SN~2005hj while the blue point is 
   SN~2012fr.}
\end{figure*}

\subsubsection{Calibration}
\label{sec:Trip_cal}
We fit equation \ref{eq:Tripp_model} (and all other models later in this
paper) using a Markov-Chain Monte Carlo (MCMC) method. 
The sampling of the MCMC chains is done using the ``No-U-Turn Sampler''
provided by the data modeling language STAN \citep{Carpenter:2017}.
Four parallel chains with
different initial positions in parameter space were produced to check for
convergence using the Gelman-Rubin statistic $\hat{R}$ \citep{GelmanRubin92}.
This statistic, which estimates the ratio of the variance of a parameter
across the 4 chains to the average variance within the chains, 
converges to $\hat{R} = 1$ typically within 500 iterations, which are 
discarded. More chains could be employed to better estimate $\hat{R}$,
but since we are only using it as a convergence test, 4 is sufficient.

The distance modulus is computed from the redshift of the
host galaxy using standard $\Lambda$CDM cosmology and a fixed 
$H_0 = 72 \kmsmpc$, 
density
parameter $\Omega_m = 0.27$, and cosmological constant parameter
$\Omega_\Lambda = 0.73$ 
to be consistent with \citet{Folatelli:2010}. 
We will introduce $H_0$ as a free parameter later in section
\ref{sec:Hubble}.  We leave the reddening parameter $R_{V}$ as a free parameter.
The error in each data point is modeled as a combination of photometric
error $\sigma_{X,i}$, intrinsic dispersion $\sigma_{X}$ for each filter $X$,
and a distance error, $\sigma_{\mu}$:
\begin{equation}
\label{eq:errors}
\sigma = \sqrt{\sigma_{X,i}^2 + \sigma_{X}^2 + \sigma_{\mu}^2}.
\end{equation}
The error $\sigma_{\mu}$ is due to peculiar velocities and is incurred when
the distance modulus is
derived from the Hubble law. Unlike $\sigma_{X}$, this extra dispersion term is
achromatic and scales with redshift: 
$\sigma_{\mu}\propto\frac{v_{pec}}{z}$, where we allow the peculiar velocity,
$v_{pec}$ to vary as a free parameter. Because the contribution of $\sigma_{\mu}$
decreases rapidly with redshift, its value will be constrained by the scatter
at low-redshift and the intrinsic dispersions $\sigma_{X}$ will be constrained
by the scatter at higher redshift. The existence of coherent flows
\citep{Neill:2007} could potentially increase $\sigma_{X}$. This will be
investigated in an upcoming CSP paper where we will leverage the increased
redshift range of the CSPII sample.

In summary, we use MCMC to fit the observed magnitudes at maximum of our SNe~
Ia by 
solving for the following parameters:  the coefficients of the polynomial
$P^N_{XYZ}$ describing the shape of the Tripp-corrected magnitudes as a
function of $s_{BV}$, the slope of the $X-Y$ color-correction $R_{XYZ}$,
the intrinsic scatter in each band $\sigma_{X}$, 
the peculiar velocity $v_{pec}$, and the slope of the luminosity-
host-mass correlation $\alpha_M$. We assume uniform priors on all parameters,
except for $\sigma_{X}$, where we impose a strictly positive uniform prior.

We investigate the effects of restricting our sample to only objects with
blue colors $(B-V)<0.5$ and also whether
we can fit the fast-declining objects ($s_{BV}<0.5$) with a single
linear or higher-order polynomial relation.

It should be noted that in cases where a magnitude in filter $X$ is corrected
by a color constructed with the same filter (e.g., $B$ corrected with $B-V$),
the appropriate error must be added to the off-diagonal elements of the covariance
matrix. This will ensure that errors are propagated properly. For example, 
degenerate cases like including both the combination $B,B,V$ and $V,B,V$,
will not improve the constraints on the model parameters any more than $B,B,V$
alone. We can also fit single $X,Y,Z$ combinations alone and obtain color
coefficients that are independent of the form of the reddening law
$R_{x}\left(R_{V}\right)$. We choose to fit a 2nd order polynomial ($N=2$) for
all color combinations. For some (e.g., $BBV$), the quadratic term is negligible
whereas in the case of the others (e.g., $YBV$), a quadratic term is significant. 
We find that,
in particular, a quadratic term is needed for the NIR bands in order to fit the
fast-declining objects.

The Tripp relations for the fits using the combinations $(X,Y,Z)=(B,B,V)$
and $(X,Y,Z)=(H,B,V)$ are shown in Figure \ref{fig:Tripp_BBV_HBV}. 
In general, we recover
many of the qualitative aspects that have been seen before
\citep{Krisciunas:2004,Folatelli:2010,Kattner:2012}, namely that both the
stretch and color corrections decrease steadily with longer wavelength.
However, we find that all the NIR stretch corrections are inconsistent with
zero slope and curvature.  In other words, even at NIR wavelengths, SNe~Ia are not 
perfect standard candles.
Table \ref{tab:Tripp_coef} shows the best-fit values for the polynomial
$P_{XYZ}^{2}$
under the same circumstances we used when fitting the intrinsic colors; namely,
omitting the fast-declining and/or reddest SNe Ia.

\subsection{Intrinsic Colors and Reddening Corrections}
\label{sec:color_reddening}
Since extinction can only make
objects redder, there will be a ``blue edge'' to the distribution of
observed colors. However, there is a strong dependence of the intrinsic
color with decline rate of SNe~Ia \citep{Phillips:1993}, so we must find
blue edges in the color-$s_{BV}$ planes (see Figure \ref{fig:color_panels}).

This was done 
in \cite{Burns:2014} by simultaneously 
solving for the color excess $E(B-V)$ and reddening slope $R_V$ of each
SN~Ia as well as the intrinsic colors (i.e., blue edges), which were
modeled as a quadratic function of $s_{BV}$.
With the increased number of objects in DR3,
particularly at low $s_{BV}$ (i.e., fast decliners), we discovered that using
a more complex fitting function was necessary in order to adequately 
fit the intrinsic colors.

For this paper, we replace the quadratic function with basis splines. This
allows for a more complex behavior of the intrinsic colors as a function of
decline rate and is simple to implement in the STAN modeling language
used to do the fits. Figure 
\ref{fig:color_panels}
shows a
comparison between these two models for the intrinsic colors. 
We use the same methodology as \citet{Burns:2014}, the only difference
being the functional form of the intrinsic colors. The largest
discrepancies between the quadratic and spline models 
are for the slowly declining ($s_{BV}>1.2$) events. Due to the
small number of objects at the slow end, one object 
(SN~2005hj) tends to pull
the solution (we have plotted this object with red points in
Figure \ref{fig:color_panels}). SN~2005hj has been shown to have a peculiar 
\ion{Si}{2}$\lambda 6355$ velocity
evolution similar to SN~2000cx \citep{Quimby:2007}, which may
explain its
peculiar colors. However, the CSP-II SN~2012fr also shows a 
similarly flat 
\ion{Si}{2}$\lambda 6355$ evolution \citep{Contreras:2018} and yet has a normal decline rate
($s_{BV} = 1.12$) and colors (plotted as blue points in 
Figure \ref{fig:color_panels}), so we do not feel justified in 
excluding it.
In Table~\ref{tab:props}, we list updated extinction values for the entire
CSP-I sample
that will be used in this paper. The details of how the basis splines are 
constructed and the values of the best-fit coefficients can be found in
Appendix~\ref{sec:Bsplines}.

% The table of SN properies with label tab:props
\capstartfalse
\begin{deluxetable*} {lcccccccccc}
\tabletypesize{\scriptsize}
\tablecolumns{11}
\tablewidth{0pt}
\tablecaption{Properties of CSPI and calibration SNe~Ia \label{tab:props}}
\tablehead{%
\colhead{SN} & \colhead{$z_{hel}$} & \colhead{$z_{cmb}$} 
& \colhead{$s_{BV}$} &
\colhead{$\Delta$ m$_{15}$($B$)} & \colhead{$m_{V,max}$} &
\colhead{$E(B-V)$} & \colhead{$R_V$} & 
\colhead{$cov(E,R)$\tablenotemark{a}} &
\colhead{$\mu_{CV}$} & \colhead{$<q_i>$}\\
\colhead{Name} & \colhead{} & \colhead{} & \colhead{} & \colhead{(mag)} &
\colhead{(mag)} & \colhead{(mag)} & \colhead{} & \colhead{} &
\colhead{(mag)} & \colhead{}}
\startdata
1981B  & $0.00603$ & $0.00717$ & 0.921(031) & 1.140(062) & 11.886(007) & 0.147(023) & 1.7(5) & $-0.0065$ & 30.89(15) & 0.98\\
1990N  & $0.00300$ & $0.00407$ & 1.122(031) & 0.961(061) & 12.677(007) & 0.128(016) & 2.3(7) & $-0.0021$ & 31.78(29) & 0.92\\
1994ae & $0.00400$ & $0.00512$ & 1.046(031) & 0.918(061) & 13.073(008) & 0.187(018) & 1.7(7) & $-0.0042$ & 32.09(26) & 0.98\\
1995al & $0.00500$ & $0.00588$ & 1.126(034) & 0.891(061) & 13.213(008) & 0.182(017) & 2.0(7) & $-0.0031$ & 32.29(23) & 0.94\\
1998aq & $0.00370$ & $0.00426$ & 0.949(030) & 1.046(060) & 12.433(006) & 0.025(012) & 2.0(3) & $-0.0008$ & 31.55(27) & 0.95\\
2001el & $0.00390$ & $0.00368$ & 0.952(031) & 1.052(061) & 12.653(006) & 0.291(016) & 2.2(3) & $-0.0029$ & 31.26(27) & 0.98\\
2002fk & $0.00712$ & $0.00661$ & 0.979(030) & 1.044(079) & 13.269(006) & 0.030(011) & 2.6(2) & $-0.0024$ & 32.45(16) & 0.98\\
2003du & $0.00638$ & $0.00665$ & 1.006(030) & 0.979(060) & 13.534(004) & 0.025(013) & 1.0(3) & $-0.0063$ & 32.84(16) & 0.98\\
2004dt & $0.01972$ & $0.01881$ & 1.189(032) & 1.104(061) & 15.068(022) & 0.149(025) & 2.7(7) & $-0.0003$ & 33.94(08) & 0.15\\
2004ef & $0.03097$ & $0.02977$ & 0.816(030) & 1.331(060) & 16.733(003) & 0.162(016) & 1.8(5) & $-0.0038$ & 35.53(05) & 0.98\\
2004eo & $0.01569$ & $0.01473$ & 0.818(030) & 1.314(060) & 15.063(008) & 0.130(025) & 1.1(7) & $-0.0098$ & 33.96(07) & 0.97\\
2004ey & $0.01578$ & $0.01463$ & 1.010(030) & 0.967(060) & 14.820(002) & 0.026(018) & 1.3(3) & $-0.0074$ & 34.12(07) & 0.97\\
2004gc & $0.03208$ & $0.03211$ & 0.923(038) & 1.079(076) & 16.576(046) & 0.226(051) & 1.9(8) & $-0.0095$ & 35.38(09) & 0.93\\
2004gs & $0.02663$ & $0.02750$ & 0.689(030) & 1.557(060) & 16.969(005) & 0.190(014) & 1.9(4) & $-0.0027$ & 35.49(05) & 0.98\\
2004gu & $0.04583$ & $0.04690$ & 1.250(032) & 0.823(061) & 17.280(009) & 0.097(031) & 1.0(7) & $-0.0048$ & 36.64(04) & 0.98\\
2005A  & $0.01913$ & $0.01834$ & 0.963(032) & 1.057(064) & 17.117(015) & 1.165(021) & 1.7(1) & $-0.0013$ & 34.37(07) & 0.97\\
2005M  & $0.02200$ & $0.02297$ & 1.208(030) & 0.796(060) & 15.907(002) & 0.055(020) & 2.0(0) & $-0.0028$ & 35.19(05) & 0.98\\
2005W  & $0.00888$ & $0.00795$ & 0.923(031) & 1.111(062) & 14.034(006) & 0.232(019) & 1.9(6) & $-0.0041$ & 32.82(15) & 0.97\\
2005ag & $0.07937$ & $0.08002$ & 1.190(031) & 0.916(060) & 18.443(005) & 0.031(018) & 1.6(3) & $-0.0024$ & 37.82(04) & 0.98\\
2005al & $0.01239$ & $0.01329$ & 0.858(030) & 1.340(060) & 14.936(005) & 0.009(009) & 1.6(3) & $-0.0019$ & 34.07(07) & 0.96\\
2005am & $0.00789$ & $0.00897$ & 0.725(030) & 1.490(060) & 13.619(004) & 0.078(016) & 1.1(8) & $-0.0060$ & 32.45(10) & 0.84\\
2005be & $0.03500$ & $0.03560$ & 0.760(035) & 1.455(073) & 16.914(039) & 0.032(022) & 1.7(3) & $-0.0019$ & 35.88(05) & 0.98\\
2005bg & $0.02307$ & $0.02416$ & 1.002(040) & 1.023(079) & 15.828(037) & 0.075(029) & 1.9(8) & $-0.0021$ & 35.04(07) & 0.98\\
2005bl & $0.02404$ & $0.02511$ & 0.387(032) & 1.906(061) & 17.806(022) & 0.330(033) & 1.9(6) & $-0.0052$ & 35.06(09) & 0.97\\
2005bo & $0.01389$ & $0.01501$ & 0.850(031) & 1.293(063) & 15.423(005) & 0.333(016) & 2.2(5) & $-0.0032$ & 33.79(10) & 0.94\\
2005cf & $0.00646$ & $0.00704$ & 0.970(031) & 1.039(102) & 13.210(007) & 0.093(019) & 2.3(6) & $-0.0062$ & 32.33(15) & 0.98\\
2005el & $0.01490$ & $0.01489$ & 0.838(030) & 1.352(060) & 14.943(007) & 0.007(008) & 1.5(4) & $-0.0031$ & 34.04(07) & 0.97\\
2005eq & $0.02896$ & $0.02835$ & 1.122(032) & 0.813(060) & 16.241(006) & 0.109(017) & 2.4(6) & $-0.0054$ & 35.40(04) & 0.98\\
2005hc & $0.04591$ & $0.04498$ & 1.193(031) & 0.875(061) & 17.305(004) & 0.037(017) & 2.4(2) & $-0.0004$ & 36.64(04) & 0.96\\
2005hj & $0.05797$ & $0.05695$ & 1.280(034) & 0.796(062) & 17.695(009) & 0.027(034) & 1.3(3) & $-0.0088$ & 37.05(04) & 0.98\\
\ldots & \ldots & \ldots & \ldots & \ldots & \ldots & \ldots & \ldots & \ldots & \ldots & \ldots \\
\enddata
\tablenotetext{a}{The covariance between $E(B-V)$ and $R_V$.}
\tablecomments{Table 2 is published in its entirety in the machine-readable
format. A portion is shown here for guidance regarding its form and content.}
\end{deluxetable*}
\capstarttrue

\subsection{The Reddening Model}
\label{sec:reddening_model}
With estimates of extinction, we can now replace the simple color term
in equation \ref{eq:Tripp_model} with a proper reddening correction.
It has been known for some time \citep{Hamuy:1995,Phillips:1993,Phillips:1999}
that the intrinsic colors of
SNe~Ia are a function of decline rate, with the fast-decliners being redder
than slow-decliners. In contrast to previous analysis \citep{Folatelli:2010},
\citet{Burns:2014} found that the relation between intrinsic colors and
$s_{BV}$ required a quadratic function rather than a linear one. It is therefore
likely that the absolute magnitudes of SNe~Ia would also be quadratic
in $s_{BV}$. We
therefore propose the following model for the observed magnitudes of our sample
of SNe~Ia:

\begin{eqnarray}
\label{eq:reddening_model}
m_{X} & = & P_{X}^{N}(s_{BV}-1)+\mu\left(z_{cmb},H_0,C\right)+ \\
      & & R_{X}\left(R_V\right)E(B-V)
      + \alpha_M\left(\log_{10}M_*/M_\odot-M_0\right). \nonumber
\end{eqnarray}
Here $m_{X}$ is the observed magnitude in filter $X$,
$P_{X}^{N}$ is an order-$N$ polynomial in $s_{BV}-1$, representing the
luminosity-decline-rate relation, $\mu\left(z_{cmb},H_0,C\right)$ is the 
distance modulus for a given redshift $z_{cmb}$,
$R_{X}\left(R_V\right)$ is the total-to-selective
absorption coefficient for filter $X$ as a function of the reddening
parameter $R_V$, and $E(B-V)$ is the color excess. The values of $R_V$ and
$E(B-V)$, as well as their errors and covariances, were computed using the
methods of \citet{Burns:2014}. 
We fit all objects in all filters simultaneously using MCMC and obtain
estimates of the following parameters:  the coefficients of the
polynomial describing the luminosity-decline-rate relation 
$P^N_{X}$, intrinsic scatter in each filter $\sigma_X$, the
slope of the host-galaxy mass relation $\alpha_M$, and the
average peculiar motion $v_{pec}$. 
As in section 
\ref{sec:Trip_cal}, we hold
the Hubble constant $H_0$ and cosmological parameters $C$ fixed and also
solve for a correlation with host galaxy mass.
In section \ref{sec:Hubble}, we will incorporate SNe~Ia with independent 
distance estimates
in order to constrain $H_0$.
When dealing with the extinction, we could correct the observed 
magnitudes $m_X$ in equation \ref{eq:reddening_model} and construct
a covariance matrix to handle the resulting correlated errors. However,
since $v_{pec}$ and $\sigma_X$ must also be included in the diagonal
terms of the covariance matrix, it would have to be re-computed and
inverted at each MCMC step. Instead, a more computationally efficient 
approach is to
treat $E(B-V)$ and $R_V$ as nuisance parameters with Gaussian
priors determined from section \ref{sec:color_reddening}. In this way,
we account for the uncertainty and covariance due to the extinction 
without having to invert large matrices.
Table \ref{tab:EBV_coef} summarizes the best-fit
parameters using several different subsets of the data.

\begin{figure*}
   \plotone{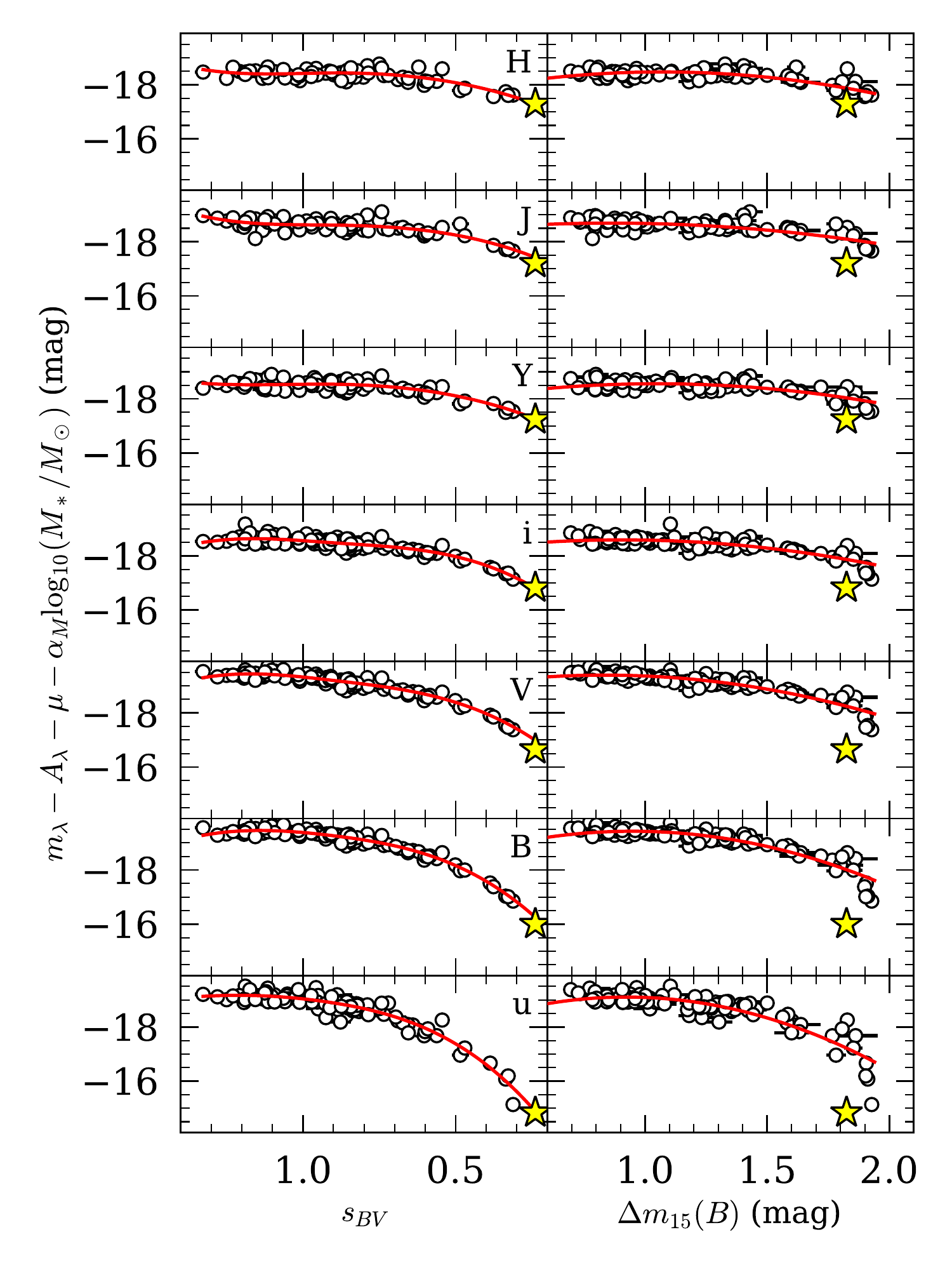}
   \caption{
      \label{fig:Phillips_relation}
      The luminosity-decline-rate relation for the CSP-I DR3 SN~Ia sample. The left-hand panels 
      show  the absolute magnitude of the  SNe~Ia as a function of 
      $s_{BV}$, whereas the right-hand panels show the absolute 
      magnitudes as a function of $\Delta m_{15}(B)$. 
      The yellow star corresponds to the fastest declining object in 
      the DR3 sample, SN~2006mr, adopting the distance of NGC~1316 
   derived from three other  normal SNe~Ia \citep{Stritzinger:2010a}.}
\end{figure*}

Figure \ref{fig:Phillips_relation} shows the results of fitting the absolute
magnitudes as a function of both $s_{BV}$ and $\Delta m_{15}(B)$. While there is a slight decrease in the RMS scatter of
the fits using $s_{BV}$ as shape parameter rather than $\Delta m_{15}$ for the
normal objects ($s_{BV} > 0.5$, $\Delta m_{15}(B) < 1.7$ mag), there is a marked
improvement in fitting the faster decliners, which appear to be a more
continuous extension of the normal (albeit quadratic) luminosity-decline-rate
relation, perhaps suggesting a single explosion mechanism 
\citep[e.g.,][]{Hoeflich:2017}. The
question then arises: can the use of $s_{BV}$ as a shape parameter allow the
use of these objects as better standardizable candles than with 
$\Delta m_{15}(B)$?

% Table of reddening-corrected luminosities with label
% tab:EBV_coef}
\capstartfalse
\begin{deluxetable*}{llllllll}
\tablewidth{0pc}
\tablecolumns{8}
\tablecaption{Absolute Magnitude Fit Parameters\label{tab:EBV_coef}}
\tablehead{
\colhead{filter} & \colhead{$P^0$} & \colhead{$P^1$} & \colhead{$P^2$} & 
   \colhead{$\alpha$} & \colhead{$\sigma_{X}$} & \colhead{$\sigma_{CV}$} & \colhead{$v_{pec}$} \\
 \colhead{} & \colhead{mag} & \colhead{mag} & \colhead{mag} & \colhead{mag/dex}
& \colhead{mag} & \colhead{mag} & \colhead{$\mathrm{km \cdot s^{-1}}$}  }
\startdata
\sidehead{Full sample}
$B$ & $-$19.422(063) & $-$0.83(11) & 3.76(28) & $-$0.084(030) & 0.11 & 0.15 & 262\\
$V$ & $-$19.352(059) & $-$0.89(10) & 2.47(26) & $-$0.072(028) & 0.10 & 0.15 & 259\\
$u$ & $-$19.106(086) & $-$1.14(16) & 5.08(39) & $-$0.175(046) & 0.17 & 0.24 & 309\\
$g$ & $-$19.448(090) & $-$0.87(11) & 3.06(28) & $-$0.092(033) & 0.11 & 0.13 & 268\\
$r$ & $-$19.216(059) & $-$0.75(10) & 2.32(24) & $-$0.081(027) & 0.10 & 0.14 & 260\\
$i$ & $-$18.589(063) & $-$0.49(11) & 1.83(26) & $-$0.083(031) & 0.12 & 0.15 & 282\\
$Y$ & $-$18.568(098) & $-$0.06(14) & 1.89(30) & $-$0.087(040) & 0.15 & 0.15 & 254\\
$J$ & $-$18.670(071) & $-$0.32(15) & 1.25(35) & $-$0.057(035) & 0.13 & 0.16 & 343\\
$H$ & $-$18.450(070) & $-$0.06(14) & 1.61(30) & $-$0.046(036) & 0.13 & 0.14 & 292\\
\sidehead{$s_{BV} > 0.5$}
$B$ & $-$19.392(061) & $-$0.89(11) & 2.76(41) & $-$0.076(029) & 0.10 & 0.16 & 283\\
$V$ & $-$19.329(059) & $-$0.93(10) & 1.73(40) & $-$0.066(028) & 0.10 & 0.14 & 275\\
$u$ & $-$19.062(086) & $-$1.25(16) & 3.62(62) & $-$0.157(045) & 0.16 & 0.23 & 283\\
$g$ & $-$19.416(084) & $-$0.93(11) & 2.07(41) & $-$0.081(031) & 0.10 & 0.13 & 302\\
$r$ & $-$19.196(060) & $-$0.78(10) & 1.72(39) & $-$0.076(027) & 0.10 & 0.14 & 267\\
$i$ & $-$18.565(064) & $-$0.53(11) & 1.08(42) & $-$0.073(031) & 0.12 & 0.15 & 273\\
$Y$ & $-$18.541(099) & $-$0.09(14) & 1.04(52) & $-$0.073(041) & 0.15 & 0.16 & 218\\
$J$ & $-$18.640(068) & $-$0.42(14) & 0.20(54) & $-$0.049(035) & 0.12 & 0.17 & 322\\
$H$ & $-$18.425(069) & $-$0.09(14) & 0.64(54) & $-$0.045(036) & 0.12 & 0.15 & 285\\
\sidehead{$E(B-V) < 0.5$}
$B$ & $-$19.404(058) & $-$0.82(10) & 3.69(26) & $-$0.079(027) & 0.10 & 0.15 & 249\\
$V$ & $-$19.335(056) & $-$0.88(09) & 2.39(23) & $-$0.067(025) & 0.09 & 0.14 & 245\\
$u$ & $-$19.096(087) & $-$1.14(16) & 5.07(40) & $-$0.175(046) & 0.17 & 0.24 & 329\\
$g$ & $-$19.429(081) & $-$0.86(10) & 2.96(26) & $-$0.083(029) & 0.09 & 0.13 & 211\\
$r$ & $-$19.202(058) & $-$0.75(09) & 2.26(22) & $-$0.076(025) & 0.09 & 0.14 & 247\\
$i$ & $-$18.577(062) & $-$0.48(10) & 1.78(24) & $-$0.083(030) & 0.12 & 0.14 & 260\\
$Y$ & $-$18.548(086) & $-$0.05(12) & 1.80(27) & $-$0.078(035) & 0.13 & 0.15 & 250\\
$J$ & $-$18.653(067) & $-$0.32(14) & 1.17(32) & $-$0.052(032) & 0.12 & 0.16 & 283\\
$H$ & $-$18.435(066) & $-$0.07(13) & 1.53(28) & $-$0.041(034) & 0.12 & 0.15 & 271\\
\sidehead{$s_{BV} > 0.5$ and $E(B-V) < 0.5$}
$B$ & $-$19.369(057) & $-$0.90(10) & 2.53(37) & $-$0.068(026) & 0.09 & 0.15 & 254\\
$V$ & $-$19.307(056) & $-$0.94(09) & 1.51(35) & $-$0.059(024) & 0.08 & 0.14 & 257\\
$u$ & $-$19.046(086) & $-$1.25(16) & 3.50(63) & $-$0.156(044) & 0.16 & 0.24 & 253\\
$g$ & $-$19.394(075) & $-$0.94(09) & 1.82(35) & $-$0.072(026) & 0.07 & 0.13 & 253\\
$r$ & $-$19.177(057) & $-$0.78(09) & 1.53(35) & $-$0.070(025) & 0.09 & 0.14 & 245\\
$i$ & $-$18.547(063) & $-$0.52(10) & 0.92(40) & $-$0.072(030) & 0.11 & 0.15 & 244\\
$Y$ & $-$18.515(087) & $-$0.10(12) & 0.77(45) & $-$0.065(035) & 0.12 & 0.16 & 232\\
$J$ & $-$18.616(065) & $-$0.43(13) & $-$0.04(48) & $-$0.042(031) & 0.11 & 0.16 & 285\\
$H$ & $-$18.403(065) & $-$0.12(13) & 0.41(50) & $-$0.039(033) & 0.11 & 0.15 & 259\\
\enddata
\tablecomments{$P^0$, $P^1$, and $P^2$ are the coefficients of the zeroth,
first, and second order terms of the polynomial $P^N_{X}\left(s_{BV}-1\right)$
from equation \ref{eq:reddening_model}.}
\end{deluxetable*}
\capstarttrue

\subsection{Results}

\begin{figure*}
\plotone{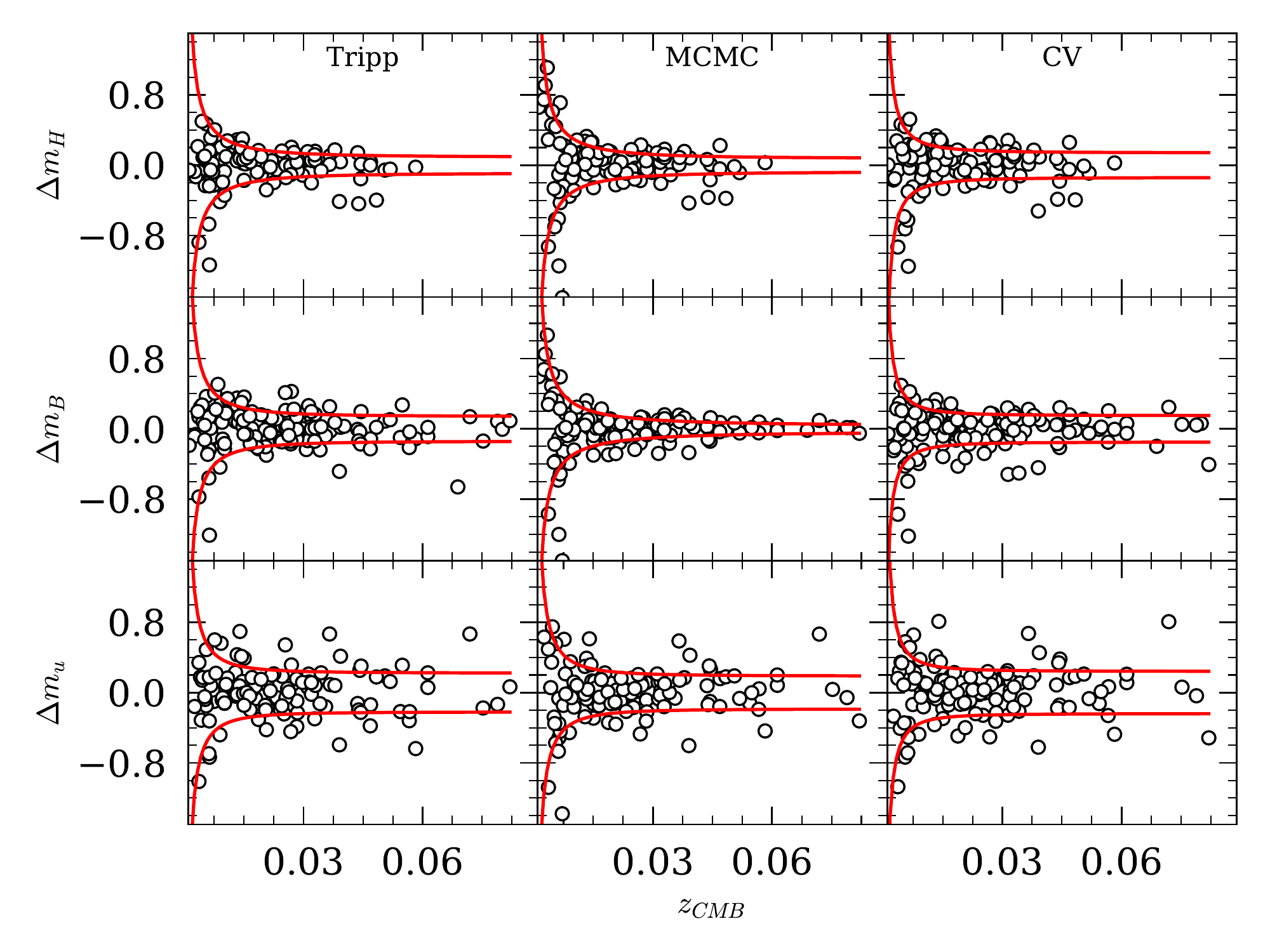}
\caption{The model residuals as a function of redshift.  The left 
panels show the residuals for a Tripp color correction, the 
middle panels for a reddening correction incorporating knowledge
of the distance, and the right panels for a reddening correction using
cross-validation where the distance is unknown. The three rows show
residuals in three different filters:  $u$, $B$, and $H$.
The solid red lines indicate the best fit for the observed dispersion,
being a combination of both peculiar velocity errors and a constant
variance for each filter.\label{fig:DM_resids}}
\end{figure*}

Using the information from our intrinsic color analysis, namely the
best-fit values of $E(B-V)$ and $R_{V}$, we can correct for the
extinction and solve for absolute luminosities rather than color-corrected
luminosities as we did in section \ref{sec:Trip_cal}. We
fit equation \ref{eq:reddening_model} for each SN Ia and each filter
$X$ simultaneously. We use the same priors for the luminosity-decline-rate
relation
coefficients, intrinsic dispersions, distance moduli, and peculiar
velocities as we did with the Tripp analysis. 
An important difference, however
is that each SN must be corrected for extinction using the values
of $E(B-V)$ and $R_V$ determined from the colors. Even though this
introduces two
additional degrees of freedom for each and every SN~Ia, there are
typically nine data points (filters) to be fit and 
the extinction parameters
were constructed without knowledge of the distances. So while it is likely
the scatter will be reduced (see Figure \ref{fig:DM_resids}), the overall
trend with $s_{BV}$ is not a function of how we fit the intrinsic
colors.

Figure \ref{fig:DM_resids} shows a comparison of the residuals of both 
the Tripp and reddening-corrected fits (left and middle panels, respectively) 
for a sample of
three filters ($u$, $B$, and $H$).
As expected, the magnitude of
these residuals increases at low redshift due to peculiar velocities
being the dominant source of variance. The red lines show the combination of
error due to peculiar velocity and an intrinsic dispersion in each filter,
whose values are tabulated in Table \ref{tab:EBV_coef}.

It is quite striking how much the dispersion is reduced in $B$ when using 
reddening-corrected magnitudes rather than a simple color correction, while
the dispersion in $u$ and $H$ seem to be unchanged. This is due to the fact
that 
while exploring parameter space, the MCMC chains will tend to favor 
values of $E(B-V)$ and $R_V$ that minimize the residuals in the 
luminosity-decline-rate relation, resulting in a posterior whose mean is
shifted with respect to that of the input prior.
These
shifts have to be small compared to the widths of their 
uncertainties and must
simultaneously improve the fit for all filters observed for each object. 
These small shifts have virtually no effect on the redder wavelengths, so
$H$ appears unchanged in Figure \ref{fig:DM_resids}. For $u$, the shifts
required to improve the fit end up increasing the scatter in the other
optical filters, and so the overall likelihood is lower, suggesting that the
scatter is real and not just errors in the reddening correction.
This is expected for $u$, where intrinsic differences from
SN-to-SN have long been known to be larger in the near UV and UV
\citep{Foley:2011b,Burns:2014} and our photometry tends to be of
poorer quality.

\begin{figure*}
\plotone{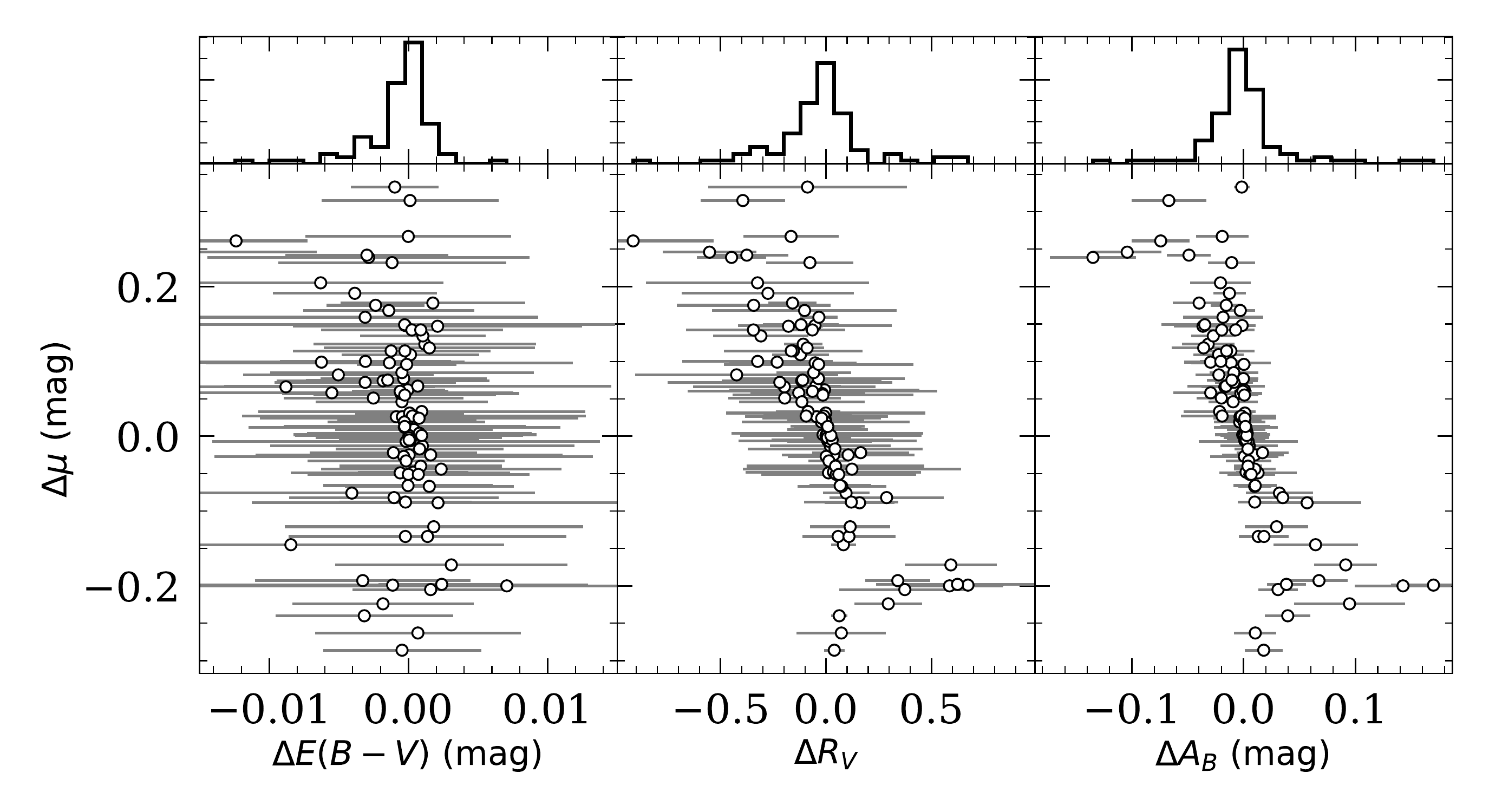}
\caption{Comparison between the distance residuals $\Delta\mu$ and the 
difference between the best-fit values of color excess (left panel),
reddening coefficient (middle panel), and total absorption in the 
$B$ band (right panel). The horizontal error bars represent the 
width of the prior that was used when calibrating the luminosity-decline-rate
relation. The top panels show histograms of the changes in
each of the extinction parameters to help show that the 
distributions are not biased.\label{fig:dDM_vars}}
\end{figure*}

\subsection{Cross Validation}
\label{sec:crossvalidate}
To more accurately measure the dispersion in the luminosity-decline-rate
relation and how minimizing the distance residuals can affect the extinction
estimates, we use the technique of cross validation, where the
calibration of the luminosity-decline-rate relation is done while omitting 
a fraction of the training sample. The resulting calibration can then be used
to predict the distance of the omitted SNe~Ia. Comparing these cross-validated
distances with the Hubble distance then gives us a more realistic measure of
the scatter one can expect when using a single SN~Ia to measure a host distance.
The cross-validated distances can also be used for the nearest
objects whose Hubble distances are uncertain due to peculiar velocities.
For simplicity, we use ``leave one out" cross-validation (LOOCV) where
the calibration is re-computed after omitting each SN in turn.

The resulting histogram of distance residuals is then fit with a Gaussian
mixture model in order to robustly measure the dispersion and also
identify outliers \citep{Hogg:2010,Krisciunas:2017}.
Briefly, the residual of object $i$ is assumed to originate from two 
Gaussian distributions: one with probability $q_i$, centered at zero and with 
standard deviation equal to the model errors (see equation \ref{eq:errors})
and a second with probability $1-q_i$ and having an unknown center and
standard deviation. The result is a more realistic estimate of $\sigma_{X}$
(which we will call $\sigma_{CV}$)
and a ``quality" parameter $q_i$ for each SN that ranges from zero to 
one. Low values of
$q_i$ indicate high probability of the SN being an outlier. The
values of $\sigma_{CV}$ are given in Table \ref{tab:EBV_coef} and 
the values of $q_i$ for each SN are given in Table \ref{tab:props}.

For each SN~Ia, we have two estimates of $E(B-V)$ and $R_V$:  those 
determined from the color analysis of section \ref{sec:color_reddening}
(when distance is not considered)
and those we get as posteriors from the MCMC while
calibrating the luminosity-decline-rate relation (where
distance is included). Figure~\ref{fig:dDM_vars}
shows how the differences in these estimates
are correlated with the residuals in the 
distance. In all cases, the shift is small compared to the width of the
prior on the variable (indicated by horizontal error-bars). Nonetheless,
the resulting change in $B$-band extinction (right panel of 
Figure~\ref{fig:dDM_vars}) is comparable to the scatter
in the luminosity-decline-rate relation and therefore will artificially reduce
the measured intrinsic scatter. Note, however, that there is no 
systematic bias:  just as many points are shifted to low values of 
$E(B-V)$ and $R_V$ as are shifted to high values.

Another important result of this cross-validation analysis is that the
dispersion in the residuals $\sigma_{CV}$ is quite uniform for all the
filters, save for $u$, which is consistently $\sim 0.1$ mag larger. Also,
combining multiple filters does not reduce the scatter relative to
using individual filters, indicating that the residuals are highly 
correlated, which is not surprising as the dominant source should be
peculiar velocities (which is an achromatic error) and uncertainties in
the reddening parameters.

We include the cross-validated measures of $\sigma_{X}$ in Table 
\ref{tab:EBV_coef}. These should be used when considering the error
in distance one can expect when using the reddening method as a distance
measure. We also include the average quality parameter $<q_i>$ and
cross-validated distance modulus $\mu_{CV}$ for each SN
in Table \ref{tab:props}.

\subsection{Fast Decliners}

We turn once again to the question of whether the fast-declining objects
for which $s<0.5$ can be incorporated into the modeling of the luminosity
of SNe Ia. Looking at Figure \ref{fig:Tripp_BBV_HBV}, it seems that
a quadratic fit to the luminosity as a function of decline-rate
is adequate to capture the behavior for $s_{BV} < 0.5$. 
A linear relation can account for the luminosity of the fast-decliners
in the $B$ band. In the case of $H$ band however, the points with 
$s_{BV} < 0.5$ lie systematically
below the linear fit and a quadratic term is required. While the 
fast-declining objects are intrinsically
dimmer than the normal SNe~Ia, they are also significantly redder
and the Tripp color correction compensates in the optical. However,
in the NIR the correction is smaller and the fast-decliners remain
below the linear relation in Figure \ref{fig:Tripp_BBV_HBV}. 
This indicates that the reason fast-declining
SNe Ia are red is likely not due to dust but rather that they are intrinsically
red.

In contrast, Figure \ref{fig:Phillips_relation} shows a very smooth and 
continuous decline-rate relation for all objects, albeit requiring a
significantly non-linear functional fit. Nevertheless it is striking that 
such a relation seems to apply to the full range of $s_{BV}$. Unfortunately,
being intrinsically faint, we have few fast-declining events that are
sufficiently distant to ascertain how well the relation does at the very
extreme end of the decline-rate relation. But for one object, SN~2006mr, which is the fastest declining object in our sample, we 
are fortunate that the host galaxy (NGC~1316) hosted 3 other SNe~Ia that
are not part of the CSP-I training sample \citep{Stritzinger:2010a}.
We therefore have an independent
distance estimate of $\mu = 31.25 \pm 0.04$ mag. This allows us to place
SN~2006mr in Figures \ref{fig:Tripp_BBV_HBV} and \ref{fig:Phillips_relation}, 
which we distinguish using
a yellow star symbol. The trend with $s_{BV}$
continues smoothly for the fastest object and SN~2006mr lies very close to
the extrapolated luminosity-decline-rate relation. More quantitatively,
we can compare the distance to NGC~1316 with the cross-validated
distance for SN~2006mr with fixed $H_0 = 72 \kmsmpc$
(to be consistent with \citep{Stritzinger:2010a}):
$\mu_{06mr} = 31.26 \pm 0.16$ mag. 
   In contrast, using $\Delta m{15}(B)$
as a predictor would lead to distance estimates up to a magnitude
more distant.

Lastly, it is worth noting that the right-hand panel of Figure
\ref{fig:Phillips_relation} shows a very similar dispersion to the left-hand
side, showing again that the true value of $s_{BV}$ primarily lies in 
how it sorts the fast-declining objects. What looks like a very fast drop-off
of the decline-rate relation for $\Delta m_{15}(B) > 1.7$ mag is in reality 
just a failure of the parameter to accurately classify how fast a
SN~Ia evolves.

\section{The Hubble Constant}
\label{sec:Hubble}
In sections \ref{sec:Trip_cal} and \ref{sec:reddening_model}
we left $H_0$ as a fixed
parameter, thereby setting the distance scale of the universe. 
In order to allow $H_0$ to vary in our simulations and infer its most 
likely value, we must use
independent distance estimates to the closest SNe~Ia. In principle, any method
can be used, but Cepheid variables have been the primary calibrator
\citep{Freedman:2001,Sandage:2006a,Riess:2016}. Cepheids have the
advantage of a long history in the literature and have well understood
systematics \citep{Madore:1991,Sandage:2006,Freedman:2010}. Their disadvantage is
that at the distance of the
closest SNe~Ia, the typical angular separations of stars in the host are small
enough to require space-based observations. Even with the 
\textit{Hubble Space Telescope} though, there is
significant crowding and overlapping point-spread functions (PSF), requiring
corrections that can approach the flux level of the Cepheid itself
\citep{Riess:2011}.

A promising alternative to Cepheids is the Tip of the Red Giant Branch (TRGB)
method \citep{Madore:2009,Jang:2017}. A significant advantage with TRGB is that
the older stellar populations being considered are found in both early- and
late-type galaxies, allowing for potentially more nearby calibrating SN~Ia
hosts. The method is also typically carried out in the outskirts of the hosts,
reducing the crowding significantly. The Carnegie-Chicago Hubble Program
(CCHP; \citealt{Beaton:2016,Freedman:2018}) aims to measure $H_0$ using population II
distance indicators and the CSP-I and CSP-II samples will be a significant component of their work.

For the purposes of this paper, we will forgo using the existing TRGB sample
as it is rather sparse and lacks SNe~Ia that were observed in the NIR. We
therefore use the Cepheid sample of \citet{Riess:2016} to calibrate our Hubble
diagram as it is the most comprehensive data set under a single 
photometric system. In the 
following sections, we present the
general method, then consider different data subsamples and their effects on
the derived value of $H_0$.

% table to hosts with cepheids and label:  tab:calibrators
\capstartfalse
\begin{deluxetable*}{lrrrrrrr}
\tabletypesize{\scriptsize}
\tablecolumns{6}
\tablecaption{Cepheid Hosts Used to Anchor Type Ia SN Distance Scale. \label{tab:calibrators}}
\tablehead{%
   \colhead{SN} & \colhead{Host} & \colhead{Opt. Ref.} & \colhead{NIR Ref.} &
   \colhead{Comments} & \colhead{$\mu$ (mag)} \\
\colhead{} & \colhead{} & \colhead{} & \colhead{} & \colhead{} & \colhead{mag} &
}
\startdata
1981B   & NGC 4536  & \citet{Buta:1983}   &  \cite{Elias:1981}             & NIR         &  30.89(06) \\
1990N   & NGC 4639  & \citet{Leibundgut:1991} &  $\cdots$                  & $\cdots$    &  31.49(08) \\
1994ae  & NGC 3370  & \citet{Riess:2005}      &  $\cdots$                  & $\cdots$    &  32.05(06) \\
1995al  & NGC 3021  & \citet{Riess:1999}      &  $\cdots$                  & $\cdots$    &  32.48(10) \\
1998aq  & NGC 3982  & \citet{Riess:2005}      &  $\cdots$                  & $\cdots$    &  31.72(08) \\
2001el  & NGC 1448  & \citet{Krisciunas:2003} & \citet{Krisciunas:2003}    & NIR         &  31.28(06) \\
2002fk  & NGC 1309  & \citet{Silverman:2012}  & \citet{Cartier:2014}       & NIR         &  32.49(07) \\
2003du  & U9391     & \citet{Hicken:2009}     & \citet{Stanishev:2007}     & NIR         &  32.88(07) \\
2005cf  & NGC 5917  & \citet{Wang:2009b}      & \citet{Friedman2015}       & NIR         &  32.25(11) \\
2007af  & NGC 5584  & \citet{Stritzinger:2011} &  \citet{Stritzinger:2011} & NIR, CSP-I  &  31.75(06) \\
2007sr  & NGC 4038  & \citet{Schweizer:2008}  &  \citet{Schweizer:2008}    & NIR, CSP-I  &  31.28(13) \\
2009ig  & NGC 1015  & \citet{Hicken:2012}     &  \citet{Friedman2015}      & NIR         &  32.47(10) \\
2011fe  & NGC 5457  & \citet{Richmond:2012} & \citet{Matheson:2012}        & NIR         &  29.16(05) \\
2011by  & NGC 3972  &  \citet{Silverman:2013}  &  \citet{Friedman2015}     & NIR         &  31.60(08) \\
2012cg  & NGC 4424  &  \citet{Marion:2016}     &  \citet{Marion:2016}      & NIR         &  31.08(32) \\
2012fr  & NGC 1365  &  \citet{Contreras:2018}  &  \citet{Contreras:2018}   & NIR, CSP-I  &  31.29(07) \\
2012ht  & NGC 3447  &  This work              &  This work                 & NIR, CSP-II &  31.88(05) \\
2013dy  & NGC 7250  &  \citet{Pan:2015}  &  \citet{Pan:2015}               & NIR         &  31.47(09) \\
2015F   & NGC 2442  &  This work            &  This work                   & NIR, CSP-II &  31.56(07) \\
\enddata
\end{deluxetable*}
\capstarttrue

\subsection{Cepheid Distances}
Table~\ref{tab:calibrators} lists the SNe~Ia we consider with Cepheid
distances from \citet{Riess:2016}, their hosts, and the source of the optical
and NIR photometry. A significant number (15/19) have SNe~Ia whose brightness
was measured in
the NIR and can be used to improve the estimates of reddening and the slope of
the reddening law (see section \ref{sec:color_reddening}). There are also 
5 SNe~Ia that
were observed by the CSP and for which there will be no systematic errors due
to differences in photometric calibration.

The models we wish to fit are the same as equations \ref{eq:Tripp_model} and
\ref{eq:reddening_model}, except that now we allow $H_0$ to vary. This will
result in a degeneracy with the $0$-th order term of $P^N_\lambda
\left(s_{BV}-1\right)$ and so we need SNe~Ia whose distances are  independent
of $H_0$. We therefore modify the distance moduli from equations
\ref{eq:Tripp_model} and \ref{eq:reddening_model} to be:
\begin{equation}
   \mu = \begin{cases}
       \mu_{ceph,i} & i \in \{\mathrm{Cepheid\ hosts}\}\\
       \mu\left(z_{hel},z_{cmb},H_0,q_0\right)& \mathrm{otherwise}.
    \end{cases}
\end{equation}
Here $\mu_{ceph,i}$ is the distance modulus of the galaxy in the set 
$\{\mathrm{Cepheid\ hosts}\}$ hosting the SN~Ia and we use the
standard second-order expansion of the luminosity distance for the
rest:
\begin{eqnarray}
\label{eq:distmod}
\mu\left(z_{hel},z_{cmb},H_0,q_0\right) & = &
     \ \ 5 \log_{10} \Bigg[\left(\frac{1+z_{hel}}
     {1+z_{cmb}}\right)\frac{c z_{cmb}}{H_0}\Big(1 + \nonumber \\
     & & \frac{1-q_0} {2}z_{cmb}\Big)\Bigg]+25
\end{eqnarray}
is the distance modulus from the Hubble law with cosmic deceleration 
$q_0 = \Omega_m/2 - \Omega_\Lambda = -0.53$
\citep{Planck-Collaboration:2016}.
The factor $(1+z_{hel})/(1+z_{cmb})$ accounts for the fact that
observational effects such as time dilation should be corrected using
redshift relative to the heliocentric frame of reference, $z_{hel}$, 
whereas cosmological distances should
be computed using redshift relative to the  Cosmic Microwave Background
(CMB), $z_{CMB}$.

The key to solving for $H_0$ therefore lies entirely in the determination of
the distances to the calibrating hosts $\mu_{ceph,i}$. These are determined 
using the Leavitt period-luminosity law with linear corrections based
on the color of the Cepheid and its metallicity \citep{Freedman:2011,Riess:2016}.
This is implemented in the following model for the observed magnitudes of
the \citet{Riess:2016} sample of Cepheid variables:

\begin{eqnarray}
   \label{eq:SN_Ceph}
   m^{Ceph}_{H} & = & M^{Ceph}_H + \mu_{ceph,i} + \alpha \log_{10}P + 
   \beta (V-I) + \nonumber \\
   & & \gamma\, [O/H].
\end{eqnarray}
Here $M^{Ceph}_H$ and $\alpha$ are the zero-point and slope of the Leavitt law,
$P$ is the period of the Cepheid, $\beta$ is the slope of the Wesenheit correction
using $V-I$ color, and $\gamma$ is the correction factor for the effect of the
metallicity $[O/H]$ \citep{Freedman:2011}. Now, we are left with a
degeneracy
between $M^{Ceph}$ and the distance moduli to the hosts $\mu_i$, which we
break by calibrating the Cepheids themselves using fundamental distance
indicators to the Large Magellanic Cloud (LMC), the water maser galaxy NGC~4258,
and galactic Cepheids with parallax measurements. We model these as follows
\begin{eqnarray}
   \label{eq:Cepheids}
   m_H^{Ceph}(LMC) & = & M^{Ceph}_H + \mu_{LMC} + \alpha \log_{10}P + \\
                   & & \beta (V-I) + \gamma [O/H]_{LMC} + \epsilon_{zp,LMC}, \nonumber\\
   m_H^{Ceph}(N4258) & = & M^{Ceph}_H + \mu_{N4258} + \alpha \log_{10}P + \nonumber \\
                    & & \beta (V-I) + \gamma [O/H] ,\nonumber \\
   m_H^{Ceph}(MW) & = & M^{Ceph}_H - 5\log_{10}\pi - 5 + \alpha \log_{10}P + \nonumber \\
                   & & \beta (V-I) + \gamma [O/H] + \epsilon_{zp,MW}, \nonumber
\end{eqnarray}
where $\mu_{LMC} = 18.49\pm 0.05$ mag \citep{Riess:2016} 
and $\mu_{N4258} = 29.40\pm 0.23$ mag \citet{Humphreys:2013} are
assigned Gaussian priors and the 10 Milky Way parallaxes $\pi$ are given by
\cite{Benedict:2007}. We include possible systematic offsets $\epsilon_{zp,LMC}$
and $\epsilon_{zp,MW}$
between the $F104W$ system used by \citet{Riess:2016} and those used by
\citet{Persson:2004} for the LMC Cepheids as well as the local Milky Way Cepheids
\citep{Groenewegen:1999}. Furthermore, we include similar terms for the
possible systematic offsets between the CSP-I natural system and the 
photometric systems listed in Table \ref{tab:calibrators}.
Since these are completely unknown, we place a
Gaussian prior on each centered at zero and with a width
equal to the error in the zero-point for each filter (see appendix
\ref{sec:zero_points}).
At this point, we do not use known Cepheids from the
DR2 release of Gaia, as they are all bright and there appears to be a 
zero-point offset in the absolute parallaxes that is dependent on 
the brightness of the star \citep{Lindegren:2018}. The systematic uncertainty
in this offset could be as high as 0.02 milli-arc-seconds (mas)
\citep{Riess:2018}, resulting
in a systematic distance error of approximately 7\%. Using Gaia as 
a robust anchor will have to await the classification and subsequent
photometric follow-up of fainter Cepheids.

When including the \citet{Riess:2016} Cepheid data, one must be very careful 
of the statistical
description of the photometry. Working in flux units with properly
weighted distributions, as we do with the SN photometry, is not possible due
to unknown bias corrections and sigma-clipping that have been performed on
the \citet{Riess:2016} data. Since the authors have not published
these corrections, one is forced to work in magnitudes. For more details,
see Appendix \ref{sec:priors}.

\begin{figure}
\plotone{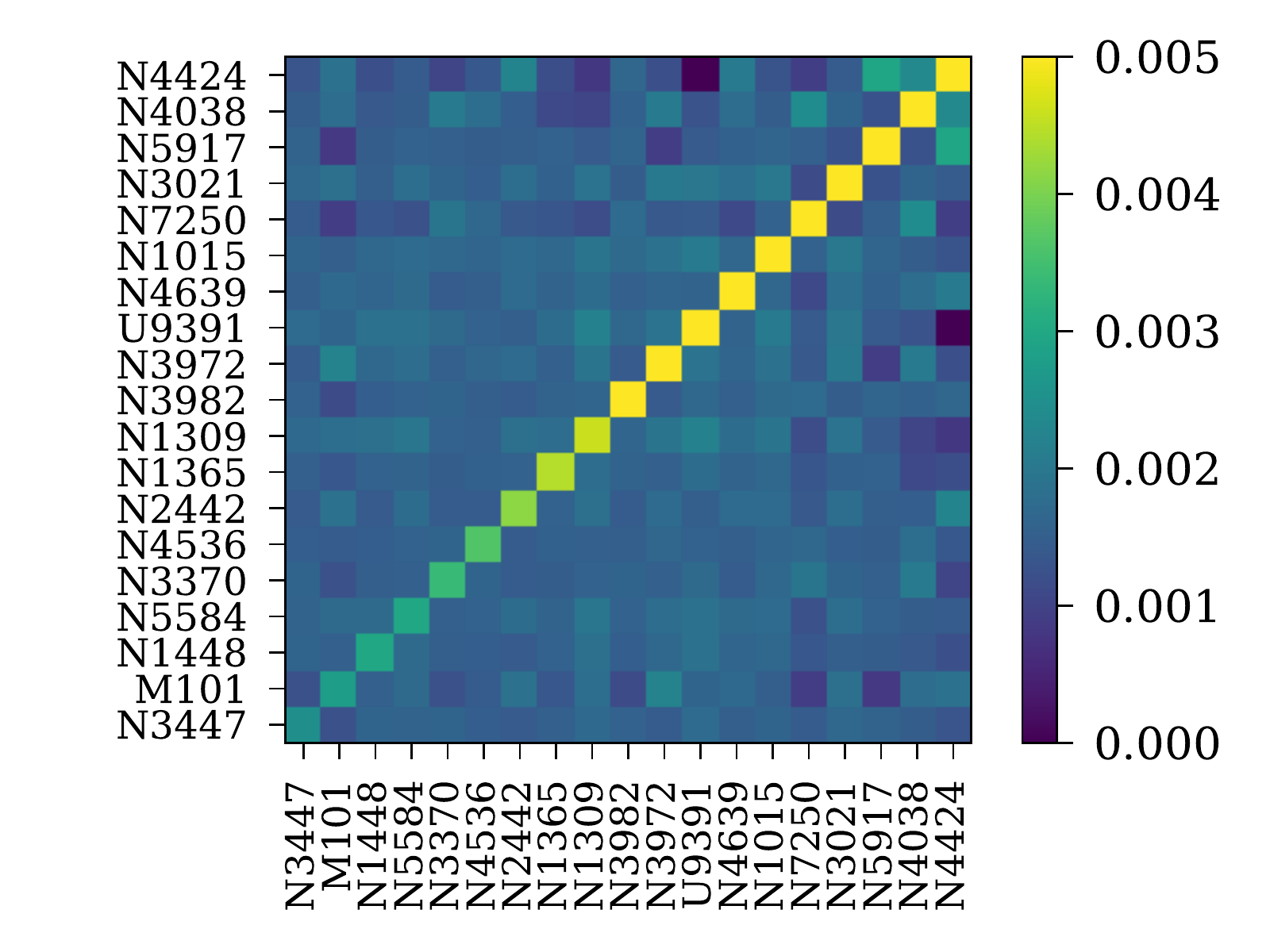}
\caption{Visualization of the covariance matrix for the calibrating host
galaxies of SNe~Ia. Each off-diagonal pixel represents the covariance between two
host galaxies while the diagonal represents the variance of a single host. The
rows and columns have been sorted by increasing variance. 
The levels have been
clipped to the interval $(0,0.005)$ in order to better visualize the
off-diagonal values.
\label{fig:covar}}
\end{figure}

Since the link between the Cepheid sample and the SN~Ia sample is the set of
distance moduli for the calibrating hosts, $\mu_{ceph,i}$, we can split the MCMC
simulation into two steps:  1) determining the values $\mu_{ceph,i}$ for
the calibrating hosts, including a complete covariance matrix $C(\mu)$, 
and 2) use these as priors for the SN~Ia MCMC runs. This allows for a much
more efficient use of computing time, as one can experiment with
how the supernova properties and priors affect $H_0$ without having to
re-compute the Cepheid calibration. We have published our covariance 
matrices $C(\mu)$ as part of the online data and they can be used by
anyone who wishes to use the host distances consistently, but not have to
deal with the Cepheid data itself. 
The software is also available to
fit with different priors and probability models.  A sample 
covariance matrix is shown in Figure \ref{fig:covar} and
shows the large range of uncertainty in the Cepheid host distances.
To better visualize the off-diagonal values, we have clipped the color
map to a maximum of $\sigma^2 = 0.005$. The true extent of 
the diagonal elements are
from an error of $\pm 0.05$ mag (NGC~3447) to $\pm 0.32$ mag (NGC~4424).
There is very little structure in the off-diagonal terms, indicating
that the primary source of covariance is the systematic error in the
distances to the fundamental anchors (LMC, NGC~4258 and MW Cepheids).

\subsection{Results}
\capstartfalse
\begin{deluxetable*}{l|llll|llll|l}
\tablewidth{0pc}
\tablecolumns{10}
\tablecaption{Best-fit values of $H_0$ in 
              $\mathrm{km\cdot s^{-1}\cdot Mpc^{-1}}$ \label{tab:H0}}
\tablehead{
 & \multicolumn{4}{c|}{Tripp} & 
\multicolumn{4}{c|}{$E(B-V)$} & \\
\colhead{filters} & \colhead{$H_0$} & \colhead{$\sigma_{Ceph}$} & 
\colhead{$\sigma_{SN}$} & \colhead{$\sigma_{total}$} &
\colhead{$H_0$} & \colhead{$\sigma_{Ceph}$} & 
\colhead{$\sigma_{SN}$} & 
\colhead{$\sigma_{total}$} & \colhead{$N_{calib}$} }
\startdata
%\multicolumn{10}{c}{Full Sample} \\
\sidehead{Full Sample}
u & $73.98$ & $1.53$ & $3.10$ & $3.45$ & $74.02$ & $1.27$ & $2.90$ & $3.17$ & 11 \\
B & $72.74$ & $1.38$ & $1.60$ & $2.11$ & $72.39$ & $1.33$ & $1.69$ & $2.15$ & 19 \\
V & $72.64$ & $1.38$ & $1.57$ & $2.09$ & $72.80$ & $1.25$ & $1.60$ & $2.03$ & 19 \\
g & $74.32$ & $1.59$ & $2.45$ & $2.92$ & $75.90$ & $1.54$ & $2.87$ & $3.26$ & 5 \\
r & $71.85$ & $1.36$ & $1.48$ & $2.01$ & $71.86$ & $1.23$ & $1.61$ & $2.02$ & 18 \\
i & $72.98$ & $1.32$ & $1.54$ & $2.03$ & $73.01$ & $1.32$ & $1.73$ & $2.18$ & 18 \\
Y & $72.25$ & $1.22$ & $2.35$ & $2.65$ & $74.28$ & $1.48$ & $3.13$ & $3.46$ & 6 \\
J & $72.47$ & $1.24$ & $1.74$ & $2.14$ & $73.11$ & $1.38$ & $2.01$ & $2.44$ & 15 \\
H & $73.86$ & $1.24$ & $1.78$ & $2.17$ & $74.54$ & $1.26$ & $2.08$ & $2.43$ & 15 \\
%YJH & $72.31$ & $1.34$ & $1.07$ & $1.72$ & $73.19$ & $1.21$ & $1.36$ & $1.81$ & 15 \\
%\multicolumn{10}{c}{$s_{BV} > 0.5$} \\
\sidehead{$s_{BV} > 0.5$}
%All & $71.35$ & $1.37$ & $0.61$ & $1.50$ & $71.90$ & $1.37$ & $0.58$ & $1.48$ & 19 \\
u & $73.75$ & $1.56$ & $3.05$ & $3.43$ & $73.37$ & $1.39$ & $2.77$ & $3.10$ & 11 \\
B & $72.44$ & $1.37$ & $1.60$ & $2.10$ & $71.79$ & $1.23$ & $1.68$ & $2.08$ & 19 \\
V & $72.42$ & $1.38$ & $1.58$ & $2.10$ & $72.33$ & $1.20$ & $1.60$ & $2.00$ & 19 \\
g & $74.10$ & $1.63$ & $2.42$ & $2.92$ & $75.28$ & $1.36$ & $2.66$ & $2.99$ & 5 \\
r & $71.61$ & $1.38$ & $1.49$ & $2.03$ & $71.50$ & $1.24$ & $1.59$ & $2.02$ & 18 \\
i & $72.78$ & $1.30$ & $1.55$ & $2.02$ & $72.73$ & $1.33$ & $1.75$ & $2.20$ & 18 \\
Y & $71.82$ & $1.23$ & $2.26$ & $2.58$ & $74.12$ & $1.57$ & $3.09$ & $3.47$ & 6 \\
J & $71.75$ & $1.20$ & $1.68$ & $2.06$ & $72.37$ & $1.21$ & $2.02$ & $2.35$ & 15 \\
H & $72.98$ & $1.17$ & $1.77$ & $2.12$ & $73.85$ & $1.13$ & $2.12$ & $2.40$ & 15 \\
%YJH & $71.47$ & $1.27$ & $1.04$ & $1.64$ & $72.60$ & $1.20$ & $1.38$ & $1.82$ & 15 \\
%\multicolumn{10}{c}{$E(B-V) < 0.5$} \\
\sidehead{$E(B-V) < 0.5$}
%All & $71.56$ & $1.38$ & $0.61$ & $1.51$ & $71.86$ & $1.36$ & $0.54$ & $1.46$ & 19 \\
u & $73.95$ & $1.55$ & $2.98$ & $3.36$ & $73.69$ & $1.17$ & $2.95$ & $3.18$ & 11 \\
B & $72.60$ & $1.40$ & $1.57$ & $2.10$ & $71.90$ & $1.18$ & $1.57$ & $1.96$ & 19 \\
V & $72.51$ & $1.36$ & $1.59$ & $2.09$ & $72.30$ & $1.25$ & $1.45$ & $1.91$ & 19 \\
g & $74.18$ & $1.54$ & $2.41$ & $2.86$ & $75.36$ & $1.47$ & $2.46$ & $2.86$ & 5 \\
r & $71.69$ & $1.38$ & $1.47$ & $2.02$ & $71.47$ & $1.27$ & $1.47$ & $1.94$ & 18 \\
i & $72.79$ & $1.35$ & $1.56$ & $2.06$ & $72.56$ & $1.31$ & $1.68$ & $2.13$ & 18 \\
Y & $72.01$ & $1.22$ & $2.27$ & $2.57$ & $73.33$ & $1.34$ & $2.67$ & $2.99$ & 6 \\
J & $71.80$ & $1.23$ & $1.71$ & $2.11$ & $72.50$ & $1.35$ & $1.84$ & $2.29$ & 15 \\
H & $73.16$ & $1.21$ & $1.77$ & $2.14$ & $74.10$ & $1.16$ & $1.99$ & $2.31$ & 15 \\
%YJH & $71.63$ & $1.27$ & $1.04$ & $1.64$ & $72.56$ & $1.23$ & $1.24$ & $1.75$ & 15 \\
\sidehead{$s_{BV} > 0.5$ and $E(B-V) < 0.5$}
%\multicolumn{10}{c}{$s_{BV} > 0.5$ and $E(B-V) < 0.5$} \\
%All & $71.57$ & $1.34$ & $0.61$ & $1.47$ & $71.36$ & $1.36$ & $0.52$ & $1.46$ & 19 \\
u & $73.99$ & $1.60$ & $2.98$ & $3.38$ & $72.93$ & $1.33$ & $2.77$ & $3.07$ & 11 \\
B & $72.67$ & $1.39$ & $1.56$ & $2.09$ & $71.19$ & $1.18$ & $1.52$ & $1.92$ & 19 \\
V & $72.52$ & $1.36$ & $1.57$ & $2.07$ & $71.70$ & $1.20$ & $1.45$ & $1.88$ & 19 \\
g & $74.11$ & $1.57$ & $2.42$ & $2.89$ & $74.67$ & $1.38$ & $2.19$ & $2.59$ & 5 \\
r & $71.73$ & $1.38$ & $1.50$ & $2.03$ & $71.03$ & $1.21$ & $1.48$ & $1.91$ & 18 \\
i & $72.77$ & $1.36$ & $1.57$ & $2.07$ & $72.16$ & $1.32$ & $1.67$ & $2.12$ & 18 \\
Y & $71.87$ & $1.25$ & $2.22$ & $2.54$ & $72.94$ & $1.40$ & $2.63$ & $2.98$ & 6 \\
J & $71.78$ & $1.32$ & $1.70$ & $2.15$ & $71.74$ & $1.25$ & $1.82$ & $2.21$ & 15 \\
H & $73.00$ & $1.18$ & $1.77$ & $2.13$ & $73.23$ & $1.08$ & $1.99$ & $2.26$ & 15 \\
%YJH & $71.54$ & $1.28$ & $1.04$ & $1.65$ & $72.23$ & $1.24$ & $1.25$ & $1.76$ & 15 \\
\enddata
\end{deluxetable*}
%\ifemulateapj
%   \capstarttrue
%\else
%   \endlongtable
%\fi

Many previous analyses have investigated numerous systematic effects
relating to the Cepheid sample, including the effects of omitting objects
based on period, metallicity, as well as the inclusion or exclusion of 
the fundamental anchors \citep[e.g.]{Riess:2016}.
Rather than run our simulations on various sub-samples 
of the Cepheid data (e.g., cut out low/high
period Cepheids, include/exclude LMC, MW, and NGC~4258, etc), we use all the
available data and do our best to model the residuals through several nuisance
parameters such as $\epsilon_{zp,i}$, intrinsic dispersions and
restricting the limits of predictor variables such as period. In this way,
we include these important systematic errors without having to run multiple
scenarios with different Cepheid sub-samples, instead focusing on the 
systematics
relating to the SNe~Ia.  The full details of the Bayesian model and
associated priors that were used can be found in Appendix \ref{sec:priors}.

\begin{figure}
   \plotone{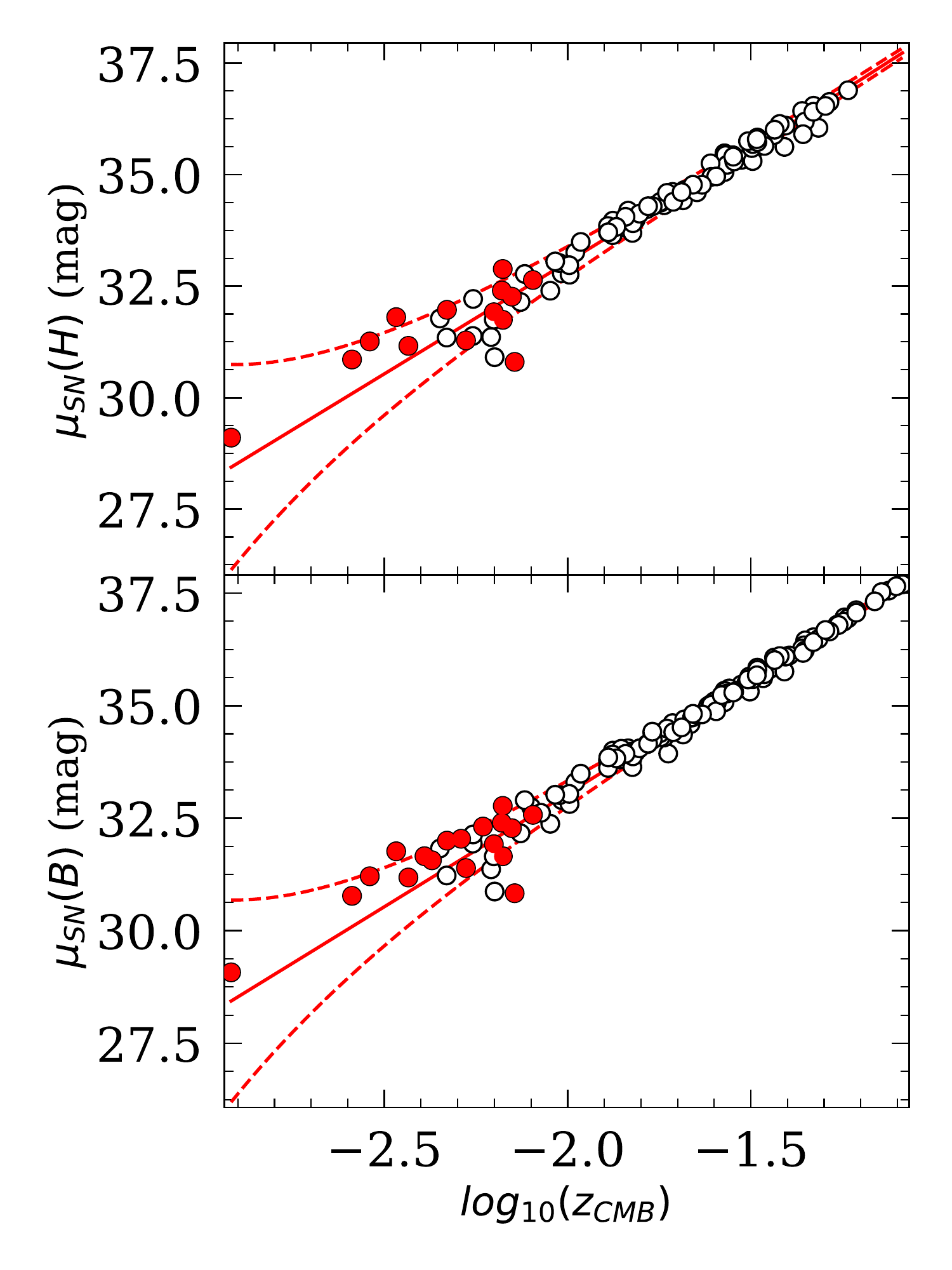}
   \caption{
      \label{fig:Hubble}
      Hubble diagram for the $H$ band (top panel) and 
      $B$ band (bottom panel) populated with SNe~Ia from the CSP-I DR3 sample, as well as with those objects
       with Cepheid hosts (plotted as solid red circles). The
      best-fit  $H_0$ = $73.5 \pm 1.5 \kmsmpc$ from combining all
      filters is shown by a solid red line.
    The expected dispersion due to intrinsic variance and peculiar velocities
   is plotted as dashed red lines. 
}
\end{figure}

Figure \ref{fig:Hubble} shows the Hubble diagram for two bands, $B$ and $H$
using the extinction-based color corrections.
The best-fit value of the Hubble constant $H_0$ is shown as solid red lines
and the predicted dispersions based and peculiar velocities and intrinsic
widths of the luminosity-decline-rate relation (labeled) are shown as dashed 
red lines. The red points correspond to the SNe in hosts with Cepheid distances.

Table \ref{tab:H0} lists a summary of the values of $H_0$ derived when using
the two different methods of dealing with the host galaxy extinction for
each of the CSP-I filters. We also split our SNe~Ia into several subsamples
as we did in sections \ref{sec:Tripp_calibration} and \ref{sec:reddening_model}.
We provide three uncertainties in $H_0$:  1) the uncertainty $\sigma_{SN}$ when
the distances to the Cepheid hosts are kept fixed, 2) the uncertainty 
$\sigma_{total}$ when all parameters are allowed to vary, and 3) the
uncertainty $\sigma_{Ceph} = \sqrt{\sigma^2_{tot} - \sigma^2_{SN}}$. This 
separation should give an indication of the error budget due to the supernova
and Cepheid data. The final column indicates the number of SNe~Ia observed
in each filter which have Cepheid distances.
The systematics
involved with the Cepheids are well-known and will not be discussed further
here. We turn instead to discussion of the SN-related effects.

The largest source of systematic error for our SN sample is the difference 
in average host galaxy stellar mass between the Cepheid sample and the more
distant sample. The limited mass range of the CSP-I sample not only 
introduces a covariance between the $H_0$ and $\alpha_M$, it also limits the
precision of our estimates for $\alpha_M$, thereby increasing the overall
systematic error in $H_0$. In principle, we could use the value of $\alpha_M$
determined from other samples as a prior, however we would be limited to
the $B$ band and we prefer to implement an independent measure of the
host mass effect. The increased range in host mass in the CSP-II sample
will greatly alleviate this systematic.

In general, the uncertainties in $H_0$ measured using the extinction
method are larger than the Tripp method. This is due to the fact that
in the Tripp method, there is a single reddening "slope" whose 
uncertainty is reduced as $\sqrt{N}$, whereas in the extinction method,
it is assumed there is an underlying distribution of $R_V$ whose width
does not decrease with SN sample size. We believe this more faithfully
describes the observed diversity in reddening properties of SN~Ia
hosts and should be included in the error budget.

A major goal of the CSP-I was to provide a Hubble constant entirely based on
NIR data (both for the Cepheids and SNe). We provide such estimates for all
three NIR bands, though clearly $Y$-band's constraint is weaker due to
it only having 5 calibrating SNe (the same is true for $g$-band). 
$J$ and $H$ have fewer calibrating SNe 
than the optical filters, but have comparable uncertainties, which is
partly due to their decreased sensitivity to both the reddening 
and host-galaxy mass corrections (see Tables \ref{tab:Tripp_coef}
and \ref{tab:EBV_coef}). In particular, $H$-band has the lowest host
mass dependence $\alpha_M = -0.04 \pm 0.03$, lowest extinction dependence,
and is nearly consistent with no color-stretch dependence. We therefore
take the $H$-band determination of $H_0 = 73.2 \pm 2.3 \kmsmpc$ as our
best estimate, a value nearly identical to that obtained by 
\citet{Riess:2016} using SNe~Ia in optical wavelengths with a Tripp
correction. Our value using Tripp and $B$-band yields 
$H_0 = 72.67 \pm 2.1 \kmsmpc$, somewhat lower, but well within
uncertainties. It is also consistent with a previous NIR determination of
the Hubble constant by \citet{Dhawan:2018}, who found
$H_0 = 72.8\pm 3.1 \kmsmpc$ and used CSP-I photometry as part of their sample.

In general, our bluer wavelengths yield estimates of
$H_0$ that are systematically lower, however we caution that at
these wavelengths, the host galaxy mass corrections are larger which,
due to the difference in average stellar mass between the distant and
Cepheid hosts, leads to  a lower value for $H_0$. Future estimates based on
CSP-II objects, which span a more representative range of
host masses, will make these estimates more reliable.

\section{Conclusions}
In this paper, we have presented an updated calibration of the CSP-I sample of
SNe~Ia using a variety of methods and assumptions and the resulting 
values for from the Hubble constant $H_0$. The 
calibrations are
most useful for the astronomical community to determine distances and
extinctions to SNe~Ia outside the CSP-I sample in order constrain host-galaxy
properties or analyze the physics of the SNe~Ia themselves.

In the forefront of our analysis is the introduction of a new light-curve shape
parameter, $s_{BV}$ that we feel is a more reliable measure of the decline rate
of SNe~Ia than $\Delta m_{15} (B)$. The reason for this is that $s_{BV}$ measures the temporal location
of a very specific and physically well-understood transition in the ejecta:
the recombination of \ion{Fe}{3} $\rightarrow$ \ion{Fe}{2}. This in turn 
depends on the temperature
evolution of the ejecta and hence the overall energy budget. In contrast,
$\Delta m_{15} (B)$ depends much more on the details of the energy transport
through the ejecta and for the fast decliners, the transition from optically
thick to thin occurs near or before day 15, leading to a break-down in the
ability of  $\Delta m_{15} (B)$ to classify these objects. Ultimately, the
validation of $s_{BV}$ as a light-curve parameter will require more
fast-declining objects in the Hubble flow. This is hampered by the fact that
the fast-decliners are also sub-luminous compared to ``normal" SNe~Ia.
Alternatively, finding fast-declining objects in galaxies that have 
red-shift independent distances, including those that have hosted
other normal SNe~Ia can also allow us to verify how precise their distances can
be estimated. This is the case with SN~2006mr, our fastest declining object.
Using the mean distance from three other SNe~Ia, we find that the distance
inferred from SN~2006mr is consistent to within the uncertainties from the
luminosity-decline-rate relation.
The ability of a single light-curve parameter to predict both
the intrinsic color and luminosity of SNe~Ia over such a large range of
decline rates suggests that there is a single explosion mechanism at work.

A great deal of effort has been done to accurately measure the intrinsic 
colors
and extinctions of our sample. This allows us to not only produce absolute
luminosities for our objects, it also allows us to deal with the fact that the
value of $R_V$ is highly variable. In the case where the extinction is high,
the colors tell us what $R_V$ is and gives a more reliable extinction
correction. In the case where extinction is low, the value of $R_V$ is
inherited from the mean of the training sample, but so is its underlying 
variance,
yielding a more reliable error in the extinction correction. We have used these
extinction estimates to produce as accurate an estimate as possible for the
extinction-corrected absolute magnitudes using photometric data on a
single well-understood system. This will be invaluable to the theoretical
community for testing explosion models 
\citep{Kasen:2007,Hsiao:2013,Stritzinger:2015,Hoeflich:2017}.

Finally, we have derived several estimates for the value of $H_0$ using two
different approaches to the color correction, multiple wavelengths,
and using different subsamples. Unsurprisingly, we find that the differences
are greatest when the average properties (decline-rate, color, and host
mass) of the calibrating SNe~Ia differ from the distant sample. In the
case of decline-rate and color, the more extreme cases in the distant
sample are rare and can be eliminated from the sample without increasing
the final error budget in $H_0$. This is not the case for host galaxy mass,
where the calibrating sample are significantly lower mass than the majority
of the distant hosts. This is simply a result of sampling the Cepheid
hosts from a smaller volume and relying on targeted searches for the
distant sample. Here, $H$-band clearly has an advantage due to its
relative insensitivity to host galaxy mass.

The use of NIR data to constrain cosmology therefore continues to show promise. 
Aside from the fact that the NIR allows one to more accurately constrain the
extinction of the SNe~Ia, its relative insensitivity to variations in 
the reddening law $R_V$ generally yields tighter constraints on $H_0$
despite having fewer Cepheid hosts. The disadvantages of the NIR are the
relative faintness of the SNe and the brightness of the background sky,
requiring more observational resources than in the optical. There is also
more uncertainty in the filter responses and zero-points
due to atmospheric effects (see Appendix \ref{sec:zero_points}),
which is particularly
worrisome for high-$z$ cosmology, where cross-band K-corrections are
required. Progress can be made here by moving to space-based observations
and/or improving our atmospheric monitoring and corrections, particularly
with regard to precipitable water vapor.

Our use of NIR data to constrain cosmology will also improve with an
increased sample of objects further out in the Hubble flow. At present,
the CSP's sample of objects has a median redshift of $z \sim 0.025$ and
are biased to high-luminosity hosts. The CSP-II, which has now finished
observations, will provide approximately 120 SNe~Ia with a median redshift
of $z \sim 0.056$ drawn from untargeted surveys. This will allow us
to more accurately determine the intrinsic dispersion of SNe~Ia and possible
correlations with host galaxy properties in the NIR.

We are also collaborating with the Carnegie Hubble Program
\citep{Freedman:2011a}, which 
seeks to establish a mid-infrared distance ladder from the Milky-way
all the way out to the hosts of SNe~Ia using Cepheids,
providing a completely independent anchor. The
CSP-II will be able to provide the final rung of this distance ladder in the
NIR out to $z \sim 0.1$.

\acknowledgments

We thank the anonymous referee for carefully reviewing this paper
and helping to improve its clarity.
We also thank Adam Riess and Dan Scolnic for their help in better understanding
the SH0ES Cepheid data.
The work of the CSP has been supported by the National Science Foundation under
grants 
AST0306969, AST0607438, AST1008343, AST1613426, and AST1613472.  M. D.
Stritzinger  acknowledges  support from the Danish Agency for Science and
Technology and Innovation through a Sapere Aude Level 2 grant and funding by a
research grant (13261) from VILLUM FONDEN.  
N. Suntzeff thanks the financial
support of the George P. and Cynthia Wood Mitchell Institute for Fundamental
Physics and Astronomy.  This paper includes data gathered with the 6.5 meter
Magellan Telescopes located at Las Campanas Observatory, Chile.  Computing
resources used for this work were made possible by a grant from the Ahmanson
Foundation.  
We thank the Cynthia and George Mitchell Foundation and Sheridan
Lorenz for their support of a number of the CSP workshops held at the Cook's
Branch Nature Conservancy where some of this work was done. This research has made use
of the NASA/IPAC Extragalactic Database (NED) which is operated by the Jet
Propulsion Laboratory, California Institute of Technology, under contract with
the National Aeronautics and Space Administration.

\newpage
\clearpage
\appendix

%\startlongtable
\capstartfalse
\begin{deluxetable}{cccc}
\tablewidth{0pc}
\tablecolumns{4}
\tablecaption{Photometry of CSP SNe~Ia in Cepheid Hosts
   \label{tab:Phot}}
\tablehead{%
\colhead{MJD (days)} & \colhead{filter} & \colhead{mag} & \colhead{phase (days)}  }
\startdata
\sidehead{SN~2012ht}
56282.3 & $B$ & 15.737(0.035) & $-$13.4\\
56283.3 & $B$ & 15.258(0.037) & $-$12.4\\
56285.3 & $B$ & 14.405(0.035) & $-$10.4\\
56286.3 & $B$ & 14.093(0.036) & $-$9.4\\
56289.3 & $B$ & 13.501(0.040) & $-$6.4\\
56290.3 & $B$ & 13.382(0.037) & $-$5.4\\
56291.3 & $B$ & 13.277(0.037) & $-$4.4\\
56292.3 & $B$ & 13.210(0.039) & $-$3.4\\
56295.3 & $B$ & 13.111(0.035) & $-$0.4\\
\ldots & \ldots & \ldots & \ldots \\
\sidehead{SN~2015F}
57092.1 & $B$ & 16.939(0.011) & $-$14.8\\
57093.1 & $B$ & 16.253(0.009) & $-$13.8\\
57094.0 & $B$ & 15.704(0.013) & $-$12.9\\
57095.1 & $B$ & 15.209(0.013) & $-$11.8\\
57096.0 & $B$ & 14.829(0.011) & $-$10.9\\
57097.1 & $B$ & 14.510(0.010) & $-$9.8\\
57098.1 & $B$ & 14.254(0.008) & $-$8.8\\
57099.0 & $B$ & 14.076(0.012) & $-$7.9\\
57100.1 & $B$ & 13.895(0.011) & $-$6.8\\
57101.0 & $B$ & 13.739(0.006) & $-$5.9\\
57102.1 & $B$ & 13.637(0.006) & $-$4.8\\
\ldots & \ldots & \ldots & \ldots \\
\enddata
\tablecomments{Table 6 is published in its entirety in the machine-readable format.  A portion is shown here for guidance regarding its form and content.}
\end{deluxetable}
\capstarttrue

\capstartfalse
\begin{deluxetable*}{ccccccccc}
\tablewidth{0pc}
\tablecolumns{9}
\tablecaption{Optical Photometry of Secondary Standards in the
standard systems of \citet{Smith:2002} and \citet{Landolt:1992}.\label{tab:LSPhot}}
\tablehead{%
   \colhead{ID} & \colhead{$\alpha$ (2000)} & \colhead{$\delta$ (2000)} & 
   \colhead{$u^\prime$} & \colhead{$g^\prime$} & \colhead{$r^\prime$} & 
   \colhead{$i^\prime$} & \colhead{$B$} & \colhead{$V$}  }
\startdata
\multicolumn{9}{c}{SN~2012ht}\\ 
 1 & 10:53:39.48 & +16:49:16.9 & 16.832(044) & 15.574(025) & 15.128(019) & 14.973(021) & 15.939(031) & 15.309(023) \\ 
 2 & 10:53:36.34 & +16:49:54.0 & 18.770(109) & 16.509(043) & 15.617(028) & 15.276(028) & 17.038(054) & 16.015(034) \\ 
 3 & 10:53:30.02 & +16:50:26.4 & 18.373(107) & 17.227(056) & 16.883(046) & 16.751(074) & 17.514(091) & 17.008(085) \\ 
 4 & 10:53:18.67 & +16:49:20.4 & 17.961(055) & 16.478(036) & 15.929(031) & 15.746(032) & 16.892(051) & 16.147(056) \\ 
 6 & 10:53:23.78 & +16:48:15.0 & $\cdots$    & 17.748(115) & 17.046(078) & 16.762(073) & 18.212(110) & 17.356(093) \\ 
 7 & 10:53:14.61 & +16:45:51.6 & $\cdots$    & 19.331(099) & 18.205(091) & 17.722(101) &  0.000(000) & 18.586(083) \\ 
 8 & 10:53:12.91 & +16:45:20.7 & $\cdots$    & 17.501(062) & 16.104(031) & 14.620(021) & 18.337(136) & 16.684(062) \\ 
10 & 10:53:19.68 & +16:43:29.5 & 19.046(037) & 18.015(117) & 17.544(054) & 17.331(108) & 18.304(111) & 17.692(087) \\ 
11 & 10:53:15.43 & +16:42:48.5 & 18.835(081) & 16.819(066) & 15.479(015) & 14.696(017) & 17.592(100) & 16.084(063) \\ 
12 & 10:53:10.80 & +16:43:24.2 & 14.540(017) & 13.371(014) & 12.991(010) & 12.874(011) & 13.693(014) & 13.139(014) \\ 
13 & 10:53:35.45 & +16:42:25.3 & 18.381(058) & 17.099(058) & 16.563(045) & 16.376(060) & 17.465(094) & 16.731(042) \\ 
\multicolumn{9}{c}{SN~2015F}\\ 
 2 & 07:35:55.06 & $-$69:24:58.4 & 15.225(005) & 13.645(003) & 13.056(006) & 12.863(006) & 14.046(011) & 13.301(003) \\ 
 3 & 07:35:57.21 & $-$69:27:12.0 & 15.181(005) & 13.864(003) & 13.373(004) & 13.196(005) & 14.212(011) & 13.575(003) \\ 
 4 & 07:35:13.04 & $-$69:29:57.7 & 15.870(005) & 14.280(003) & 13.612(003) & 13.333(004) & 14.709(012) & 13.903(003) \\ 
 5 & 07:35:02.17 & $-$69:23:12.0 & 16.885(008) & 14.492(003) & 13.532(004) & 13.134(023) & 15.081(014) & 13.953(003) \\ 
 6 & 07:36:01.21 & $-$69:37:45.1 & 15.753(006) & 14.358(003) & 13.823(003) & 13.621(003) & 14.729(010) & 14.051(004) \\ 
 7 & 07:36:57.82 & $-$69:36:11.5 & 16.231(005) & 14.570(003) & 13.876(002) & 13.603(003) & 15.017(012) & 14.184(003) \\ 
 8 & 07:35:32.74 & $-$69:34:20.3 & 15.969(005) & 14.594(003) & 14.048(002) & 13.834(003) & 14.967(011) & 14.278(003) \\ 
 9 & 07:37:27.92 & $-$69:26:45.3 & 16.142(005) & 14.668(003) & 14.118(002) & 13.924(002) & 15.057(013) & 14.353(003) \\ 
10 & 07:34:59.88 & $-$69:35:36.1 & 17.677(010) & 14.996(003) & 13.944(002) & 13.488(003) & 15.610(017) & 14.418(003) \\ 
11 & 07:37:26.81 & $-$69:27:13.5 & 16.503(006) & 14.819(003) & 14.174(002) & 13.942(002) & 15.254(009) & 14.452(003) \\ 
12 & 07:35:13.96 & $-$69:23:25.6 & 16.250(005) & 14.833(003) & 14.253(002) & 14.023(002) & 15.212(016) & 14.496(003) \\ 
13 & 07:35:54.32 & $-$69:24:19.3 & 16.882(007) & 14.997(003) & 14.218(002) & 13.916(002) & 15.485(023) & 14.552(003) \\ 
14 & 07:37:09.56 & $-$69:26:09.1 & 16.201(005) & 14.938(003) & 14.472(002) & 14.291(002) & 15.275(013) & 14.667(003) \\ 
15 & 07:35:06.59 & $-$69:35:23.9 & 16.842(007) & 15.082(003) & 14.379(002) & 14.094(002) & 15.534(017) & 14.685(003) \\ 
16 & 07:36:26.34 & $-$69:26:17.8 & 16.293(006) & 15.045(003) & 14.574(002) & 14.399(002) & 15.380(016) & 14.770(003) \\ 
17 & 07:37:21.60 & $-$69:26:51.5 & 16.593(006) & 15.138(003) & 14.577(002) & 14.361(002) & 15.535(010) & 14.817(003) \\ 
18 & 07:36:31.03 & $-$69:29:30.6 & 16.707(006) & 15.225(003) & 14.563(002) & 14.287(002) & 15.641(019) & 14.860(003) \\ 
19 & 07:36:23.37 & $-$69:23:58.9 & 17.164(008) & 15.367(003) & 14.637(002) & 14.354(002) & 15.828(027) & 14.953(003) \\ 
20 & 07:36:31.45 & $-$69:37:16.7 & 16.816(007) & 15.330(003) & 14.730(003) & 14.497(002) & 15.736(014) & 14.989(003) \\ 
21 & 07:37:24.84 & $-$69:30:31.9 & 16.719(006) & 15.345(003) & 14.763(003) & 14.524(002) & 15.728(016) & 15.016(003) \\ 
22 & 07:37:02.86 & $-$69:32:36.1 & 17.664(010) & 15.516(003) & 14.694(003) & 14.399(002) & 16.018(020) & 15.062(003) \\ 
23 & 07:35:37.73 & $-$69:27:52.9 & 17.065(007) & 15.545(003) & 14.910(003) & 14.684(002) & 15.947(022) & 15.177(003) \\ 
24 & 07:36:48.87 & $-$69:26:04.5 & 16.864(007) & 15.553(003) & 15.027(003) & 14.824(002) & 15.913(024) & 15.252(003) \\ 
25 & 07:37:15.38 & $-$69:26:48.0 & 16.876(007) & 15.565(003) & 15.063(003) & 14.880(002) & 15.925(018) & 15.273(003) \\ 
26 & 07:35:07.75 & $-$69:23:12.9 & 16.883(008) & 15.376(003) & 14.741(003) & 14.495(003) & 15.776(022) & 15.009(003) \\ 
27 & 07:36:47.77 & $-$69:35:48.1 & 17.222(008) & 15.769(003) & 15.145(003) & 14.900(002) & 16.186(022) & 15.411(003) \\ 
28 & 07:35:44.55 & $-$69:36:21.2 & 17.429(009) & 15.824(003) & 15.121(003) & 14.826(003) & 16.256(025) & 15.435(003) \\ 
29 & 07:36:34.68 & $-$69:29:35.7 & 18.523(019) & 16.037(003) & 14.978(003) & 14.564(002) & 16.589(024) & 15.481(003) \\ 
30 & 07:36:41.69 & $-$69:24:00.0 & 17.420(009) & 15.915(003) & 15.289(003) & 15.024(003) & 16.323(028) & 15.562(003) \\ 
31 & 07:37:07.50 & $-$69:33:46.2 & 17.587(010) & 15.964(003) & 15.336(003) & 15.113(003) & 16.396(022) & 15.605(003) \\ 
\enddata
\end{deluxetable*}
\capstarttrue

\capstartfalse
\begin{deluxetable*}{cccccc}
\tablewidth{0pc}
\tablecolumns{6}
\tablecaption{NIR Photometry of Secondary Standards\label{tab:LSNIRPhot}}
\tablehead{%
   \colhead{ID} & \colhead{$\alpha$ (2000)} & \colhead{$\delta$ (2000)} & 
\colhead{$Y$} & \colhead{$J$} & \colhead{$H$} }
\startdata
\multicolumn{6}{c}{SN~2012ht}\\ 
101 & 10:53:23.91 & +16:46:38.9 & 15.088(030) & 14.608(021) & 13.954(045)  \\ 
102 & 10:53:21.20 & +16:48:00.0 & 15.601(039) & 15.149(011) & 14.556(037)  \\ 
103 & 10:53:23.81 & +16:48:14.9 & 15.922(026) & 15.627(027) & 15.114(013)  \\ 
104 & 10:53:28.19 & +16:46:49.9 & 16.512(087) & 16.256(095) & 15.501(041)  \\ 
105 & 10:53:17.48 & +16:45:05.1 & 16.807(062) & 16.371(012) & 15.841(084)  \\ 
106 & 10:53:15.49 & +16:46:55.3 & 16.989(054) & 16.407(031) & 15.902(063)  \\ 
107 & 10:53:21.01 & +16:45:22.0 & 17.919(103) & 17.414(055) & 16.884(194)  \\ 
\multicolumn{6}{c}{SN~2015F}\\ 
101 & 07:36:31.00 & $-$69:29:31.1 & 13.494(024) & 13.162(023) & 12.723(028)  \\ 
102 & 07:36:34.61 & $-$69:29:36.3 & 13.636(026) & 13.239(027) & 12.645(039)  \\ 
103 & 07:36:07.16 & $-$69:28:51.8 & 14.796(047) & 14.449(062) & 14.052(069)  \\ 
104 & 07:36:31.16 & $-$69:29:59.5 & 15.158(044) & 14.898(038) & 14.557(060)  \\ 
106 & 07:36:17.75 & $-$69:28:28.8 & 15.755(048) & 15.236(042) & 14.605(139)  \\ 
107 & 07:36:22.93 & $-$69:30:25.6 & 15.652(063) & 15.264(044) & 14.759(045)  \\ 
108 & 07:36:18.43 & $-$69:28:59.9 & 16.110(059) & 15.713(052) & 15.139(066)  \\ 
109 & 07:36:15.17 & $-$69:29:36.7 & 16.709(055) & 16.232(085) & 15.575(143)  \\ 
110 & 07:36:22.02 & $-$69:28:50.5 & 16.798(097) & 16.344(120) & 15.773(132)  \\ 
111 & 07:36:32.03 & $-$69:28:59.4 & 16.822(067) & 16.531(155) & 16.103(128)  \\ 
112 & 07:36:01.57 & $-$69:30:57.5 & 16.951(054) & 16.563(159) & 15.736(269)  \\ 
113 & 07:36:36.41 & $-$69:29:10.4 &  0.000(000) & 16.432(079) & 16.066(076)  \\ 
114 & 07:36:08.92 & $-$69:30:22.5 & 16.924(128) & 16.475(124) & 15.715(222)  \\ 
115 & 07:36:13.37 & $-$69:29:41.5 & 17.362(088) & 16.890(130) & 16.324(106)  \\ 
116 & 07:36:02.64 & $-$69:28:41.1 & 17.375(093) & 16.758(089) & 15.867(090)  \\ 
117 & 07:36:36.42 & $-$69:29:56.9 &  0.000(000) & 16.432(081) & 16.576(195)  \\ 
118 & 07:36:20.70 & $-$69:30:09.5 & 17.462(097) & 17.037(132) & $\cdots$     \\ 
119 & 07:36:01.19 & $-$69:29:02.9 & 17.358(408) & 16.865(420) & 15.706(103)  \\ 
120 & 07:36:22.39 & $-$69:31:12.2 & 17.313(141) & 16.722(132) & 15.912(126)  \\ 
121 & 07:36:33.40 & $-$69:29:07.7 & 17.639(091) & 17.031(123) & 16.429(101)  \\ 
122 & 07:36:06.88 & $-$69:30:25.3 & 17.253(025) & 16.767(050) & 15.952(097)  \\ 
123 & 07:36:29.68 & $-$69:29:22.3 & $\cdots$    & 17.232(130) & 16.345(090)  \\ 
\enddata
\end{deluxetable*}
\capstarttrue

\section{A. Photometry of SN~2012ht and SN~2015F}
\label{sec:12ht+15F}
In section \ref{sec:Hubble} we used two SNe~Ia from the CSP-II 
project 
to anchor the SN~Ia distance ladder and so present their photometry in 
this section. The CSP-II is a continuation
of CSP-I, with a particular emphasis
on the NIR observations at higher red-shift than in the CSP-I. The observational
setup and 
procedure in the optical is identical to CSP-I and details are given by
\citet{Krisciunas:2017}. For the NIR observations, we moved RetroCam from the
Swope telescope to the duPont at LCO. Other than that, our observational 
procedures and data reduction are
identical to the CSP-I and the complete telescope, filter, and CCD transmission 
functions have
been measured, which are available at the CSP website\footnote{
\href{https://csp.obs.carnegiescience.edu}{https://csp.obs.carnegiescience.edu}}. 
Note that in October of 2013, the CCD detector on the Swope telescope was 
upgraded from a Site3 to e2v CCD. This resulted
in a change to our filter functions and zero-points in the optical. SN~2012ht is
therefore on the old CSP-I natural system, whereas SN~2015F is on the new one.
Both SNe are on the new duPont RetroCam natural system, which is described
in \citet{Contreras:2018}.

Table~\ref{tab:Phot} lists the photometry of SN~2012ht and SN~2015F. 
Tables \ref{tab:LSPhot} and \ref{tab:LSNIRPhot}
list the photometry of the reference stars in the standard 
optical \citep{Landolt:1992,Smith:2002} and NIR \citep{Persson:1998} systems.
The filter functions and photometric zero-points $zp_\lambda$
of the CSP-I and CSP-II
natural systems are available at the CSP website.
These can be used to S-correct \citep{Stritzinger:2005} the photometry to 
other systems (see appendix \ref{sec:zero_points}).

\begin{figure*}
   \plotone{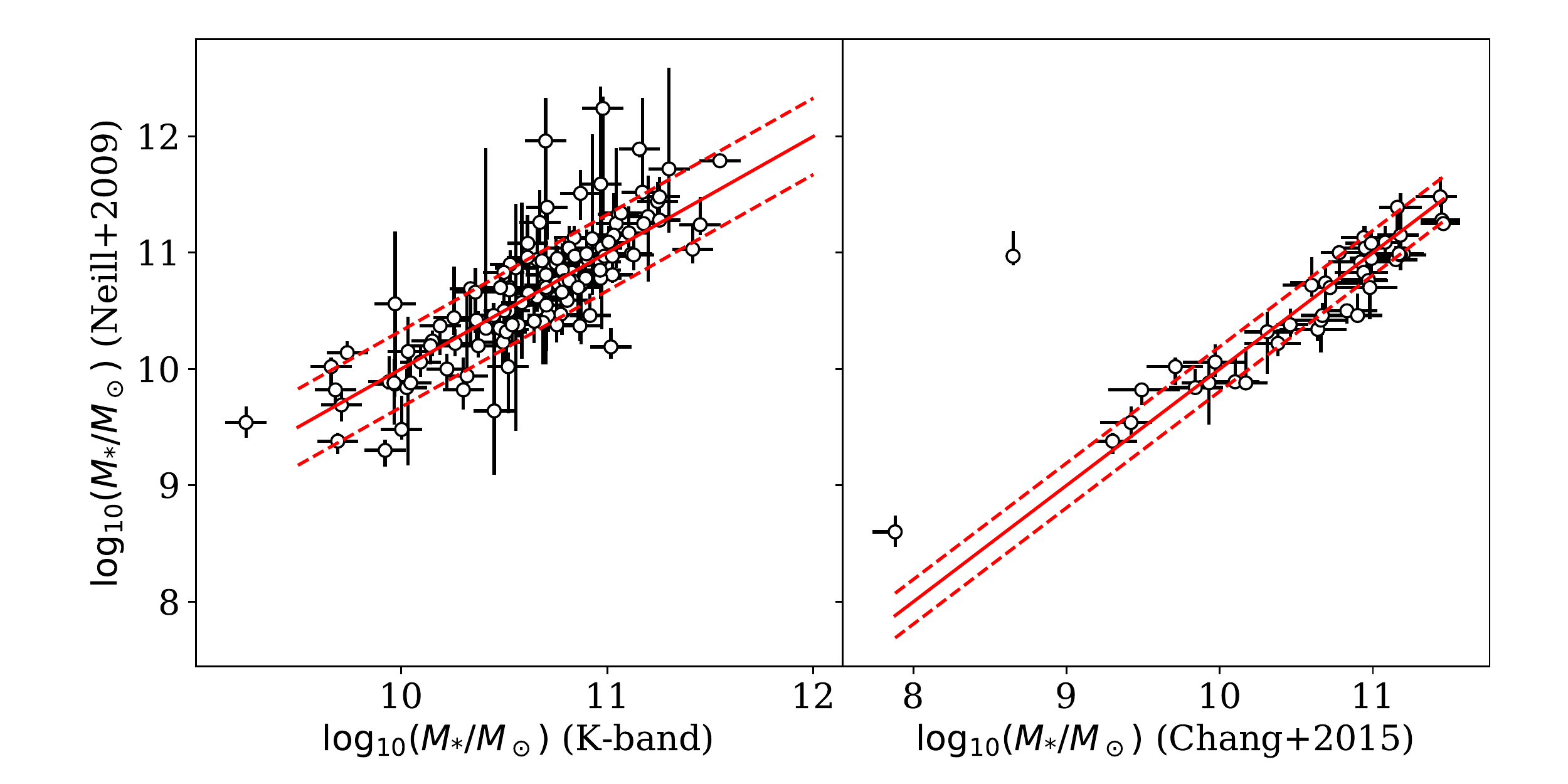}
   \caption{Comparison of host galaxy stellar masses determined from
      2MASS $K$-band photometry (left panel) and
      masses from \citet{Chang:2015} (right panel) with those from 
      \citet{Neill:2009}. The
      solid red lines are fits assuming a slope of 1 and the
      dashed lines indicate $\pm 0.3$ dex.
   \label{fig:Mstellar_comp}}
\end{figure*}

\section{B. Host Galaxy Mass}
\label{sec:HostMass}
In recent years, there has been evidence of a correlation between the
color- and decline-rate-corrected luminosity of SNe~Ia and bulk 
properties of their host galaxies. For nearby SNe~Ia, \citet{Neill:2009} 
found a correlation
between host age and corrected peak brightness and \citet{Kelly:2010} found a 
2.5-$\sigma$
correlation between host galaxy stellar mass and corrected peak brightness,
both using
a sample similar to the CSP-I sample. Using a more distant
sample from the Supernova Legacy Survey (SNLS), \citet{Sullivan:2010}
show a similar trend with stellar mass, measuring a non-zero gradient
at $\sim 3\sigma$. More recently, \citet{Uddin:2017} have used a comprehensive
set of more than 1300 SNe~Ia and detect a correlation between Hubble
residual and host galaxy mass at a significance of $4\sigma$.

This effect can bias our determination of the Hubble constant if the sample
of SNe~Ia in CSP-I have hosts with significantly different stellar mass
than the 19 hosts with Cepheid distances used to anchor the Hubble diagram.
To investigate this, we determine stellar masses for a sub-sample of CSP-I
hosts using the 2MASS extended source catalog \citep{Jarrett:2000}. We then
assume a constant mass-to-light ratio in K-band \citep{McGaugh:2014}.
Under this assumption, the stellar mass of a host galaxy is given by:
\begin{equation}
   \log_{10}\left(M_{*}/M_\odot\right) = -0.4 \left(m_K - \mu\right) + C
   \label{eq:stellar_mass}
\end{equation}
where $m_K$ is the apparent $K$ magnitude of the host, $\mu$ is its
distance modulus, and
$C$ is a constant which sets the mass scale. To determine $C$, we apply 
equation \ref{eq:stellar_mass} to the sample of galaxies from 
\citet{Neill:2009} which have 2MASS $K$-band photometry. The
left panel of Figure 
\ref{fig:Mstellar_comp} shows the comparison of the two estimates in
stellar mass. The best-fit value for the mass scale is $C = -1.04$ dex and
the RMS scatter is $\pm 0.3$ dex.

Of the 120 SNe~Ia in the CSP-I sample, 103 have 2MASS measurements of the
host galaxy and their $K$-band magnitudes and corresponding stellar masses
are listed in table \ref{tab:props}.
An additional four objects (SN~2003du, SN~2005ir, SN~2006ej, and SN~2008bf),
are in the \citet{Neill:2009} sample and can be used directly. 
Lastly, we use host mass estimates from \citet{Chang:2015} for another
8 objects, including the host of SN~2012ht. To check for consistency, 
we have plotted the \citet{Neill:2009} masses versus the \citet{Chang:2015}
masses for 46 objects they have in common. Aside from one clear outlier,
the correspondence is very good with an RMS scatter of 0.2 dex. 
This leaves us with only 
5 objects for which we have no host mass estimates. These are left as
free parameters with uniform priors over the range of stellar masses
observed for the CSP-I sample: $9 < \log_{10}\left(M_*/M_\odot\right) < 11.5$.

\begin{figure*}
   \plotone{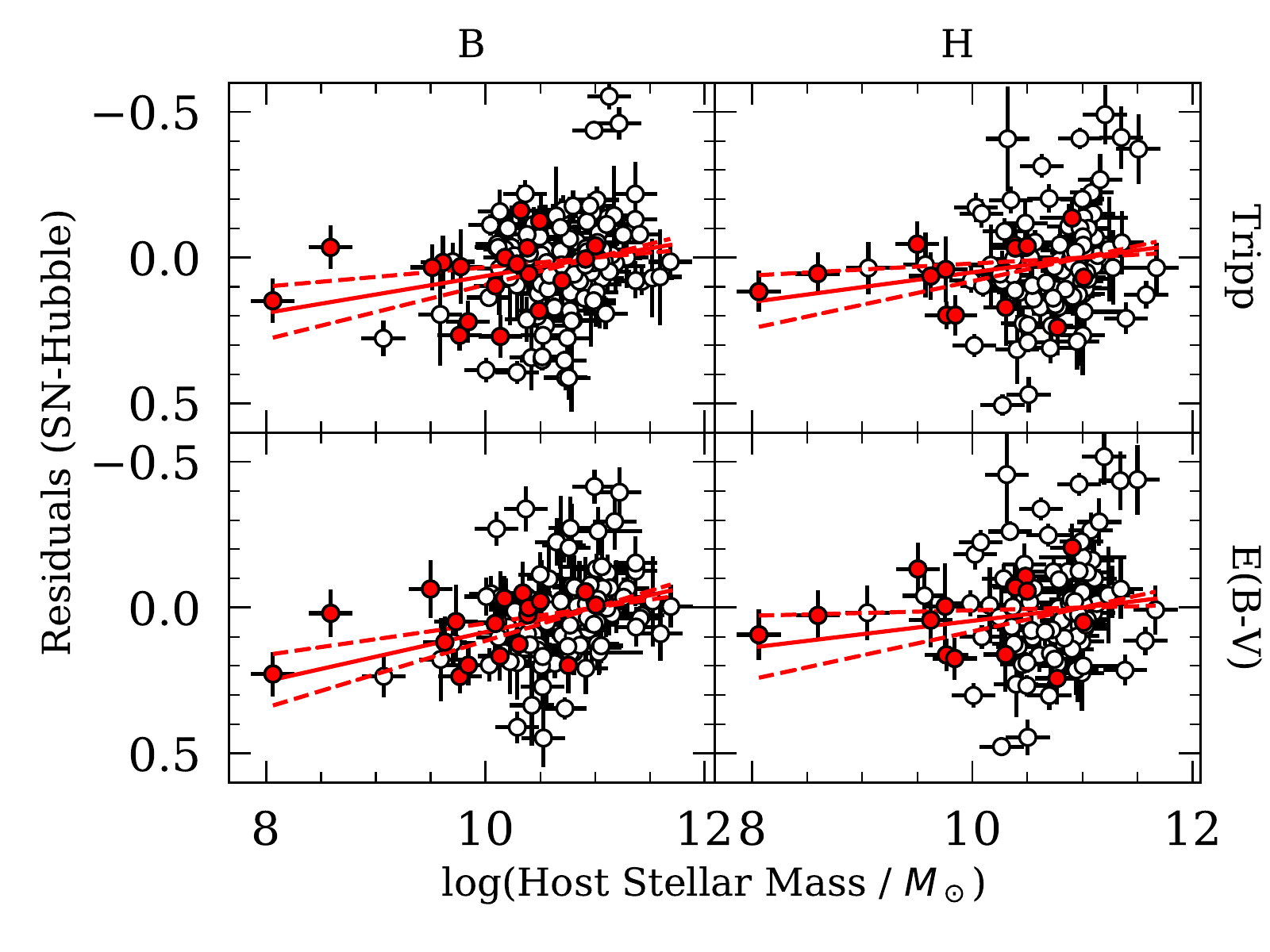}
   \caption{The residuals in the Hubble diagram as a function of stellar
      mass of the host galaxies. The top row of panels shows residuals from
      the Tripp model, while the bottom row shows residuals from the 
      extinction method. The left column is for $B$-band, while the right
      column is $H$-band. The best-fit lines are shown as red solid lines
      with the 1-$\sigma$ as dashed red lines. The SNe~Ia with Cepheid
      distances are plotted as red points.\label{fig:res_Msun}}
\end{figure*}

The mean stellar mass of
the CSP-I sample excluding the Cepheid hosts is 
$\log_{10}\left(M_*/M_\odot\right) = 10.7$ whereas the mean stellar mass
of the Cepheid hosts is $\log_{10}\left(M_*/M_\odot\right) = 10.1$, or
a difference of $0.6$ dex. This 
is large enough to produce a 2-3\% shift in $H_0$,
given the typical host mass-luminosity slopes that are measured. 
We therefore
include a linear correction factor in equations \ref{eq:Tripp_model}
and \ref{eq:reddening_model}. The slopes of the host mass corrections
are given in tables \ref{tab:Tripp_coef} and \ref{tab:EBV_coef}.
Figure \ref{fig:res_Msun} shows the correlation between Hubble residuals
and host mass using two different filters ($B$ and $H$) and the two
different methods of treating extinction. In all cases, the slope is
significant to between 1 and 2-$\sigma$, but generally decreasing
with wavelength. The hosts with Cepheid distances
are colored with red points.

\capstartfalse
\begin{deluxetable*}{llll}
%\tabletypesize{\scriptsize}
\tablecolumns{4}
%\tablewidth{0pt}
\tablecaption{Bayesian Model Parameters\label{tab:parameters}}
\tablehead{
\colhead{Parameter} & \colhead{Description} & \colhead{Prior} & \colhead{Typical Value} }
\startdata
\multicolumn{4}{c}{SNe~Ia} \\
$H_0$ & Hubble constant & $U(-\infty,\infty)$ & see Table~\ref{tab:H0} \\
$P^N_\lambda$ & Coefficients of the decline-rate relation& 
   $U(-\infty,\infty)$ & see Tables~\ref{tab:Tripp_coef} and 
   \ref{tab:EBV_coef}\\
$R_{XYZ}$ & Tripp coefficient for filter $X$ corrected by color $Y-Z$ &
   $U(0,\infty)$ & see Table \ref{tab:Tripp_coef} \\
$[E(B-V)_i,R_{V,i}]$ & Color excess and reddening slope & 
           see \citep{Burns:2014} & see Table~\ref{tab:props} \\
$\alpha$ & Slope of the host galaxy mass-luminosity correction & 
   $U(-\infty,\infty)$ & see Tables~\ref{tab:Tripp_coef} and
      \ref{tab:EBV_coef}\\
$v_{pec}$ & Peculiar velocity of SN hosts & 
           $U(0,\infty)$ & $200 - 300\ \mathrm{km\cdot s^{-1}}$ \\ 
$\sigma_\lambda$ & Intrinsic dispersion in filter $\lambda$ & 
           $U(0,\infty)$ & see Tables \ref{tab:Tripp_coef} and 
           \ref{tab:EBV_coef}\\
$q_i$ & Outlier probability of data point $i$ & U(0,1) & See Table \ref{tab:props} \\
\hline
\multicolumn{4}{c}{Cepheids} \\
$M_{ceph,H}$ & Cepheid absolute magnitude zero-point & 
               $U(-\infty,\infty)$ & $-2.59(09)$ mag\\ 
$\alpha$ & Slope of the period-luminosity relation & 
               $U(-\infty,\infty)$ & $-3.24(04)$ mag\\ 
$\beta$ & Slope of the period-color relation & 
               $U(-\infty,\infty)$ & $0.30(05)$ \\ 
$\gamma$ & Slope of the metallicity-luminosity relation & 
               $U(-\infty,\infty)$ & $-0.07(09)$ mag/dex\\ 
$\sigma_{ceph}$ & intrinsic dispersion & 
               $U(0,\infty)$ & $0.30(01)$\\ 
$\sigma_{OH}$ & Uncertainty in $[O/H]$ measurements & $U(0,\infty)$ &
               $0.78(66)$ dex\\
$P_{min},P_{max}$ & Period limits for Cepheid sample &
               $U(0,\infty)$ & $5.10(06)$,$98.78(04)$ days\\ 
$q_i$ & Mixture fraction for data point $i$ & U(0,1) & 0.65 - 0.80 \\
$m_c$ & Mixture offset for 2nd component & $U(-\infty,\infty)$ & 
               $0.14$ - $0.25$ mag\\
$\sigma_c$ & Mixture width for 2nd component & $U(0,\infty)$ & 
               $0.25$ - $0.71$mag\\
\enddata
\end{deluxetable*}
\capstarttrue

\section{C. Intrinsic Color Model with Bsplines}
\label{sec:Bsplines}
\citet{Burns:2014} modeled the intrinsic color as a second-order
polynomial, which therefore had three degrees of freedom per filter. 
There is no physical basis for this and it was 
chosen 
primarily as a numerical convenience. Any functional form that captures the 
shape of the ``blue edge'' in Figure \ref{fig:color_panels} will suffice. With 
additional data from CSP-I, it became apparent that a simple polynomial was
insufficient. Basis splines provide more flexibility needed to fit the complex
behavior. Also, the sample of objects is sparse at either end of the
luminosity-decline-rate relation and individual objects 
at either end of the distribution in $s_{BV}$ will not influence 
the overall shape
of a spline  as much as it would for a 
polynomial.

Basis splines are constructed using basis functions:
\begin{equation}
\label{eq:Bsplines}
S\left(s_{BV}\right) = \sum_i a_i B_i\left(s_{BV}\right).
\end{equation}
Here $S$ is the spline function used to model the intrinsic colors,
$a_i$ are the spline coefficients controlling the shape, and
$B_i$ are the basis functions, which are
constructed recursively using the Cox-de Boor algorithm 
\citep{deBoor:1978}. This algorithm is 
available as a standard library for most scientific computing languages. 
The values of $B_i\left(s_{BV}\right)$ can be computed for each SN~Ia
and then passed to STAN as data, which will solve for the $a_i$ as
free parameters of the model. 

\begin{figure}
\plotone{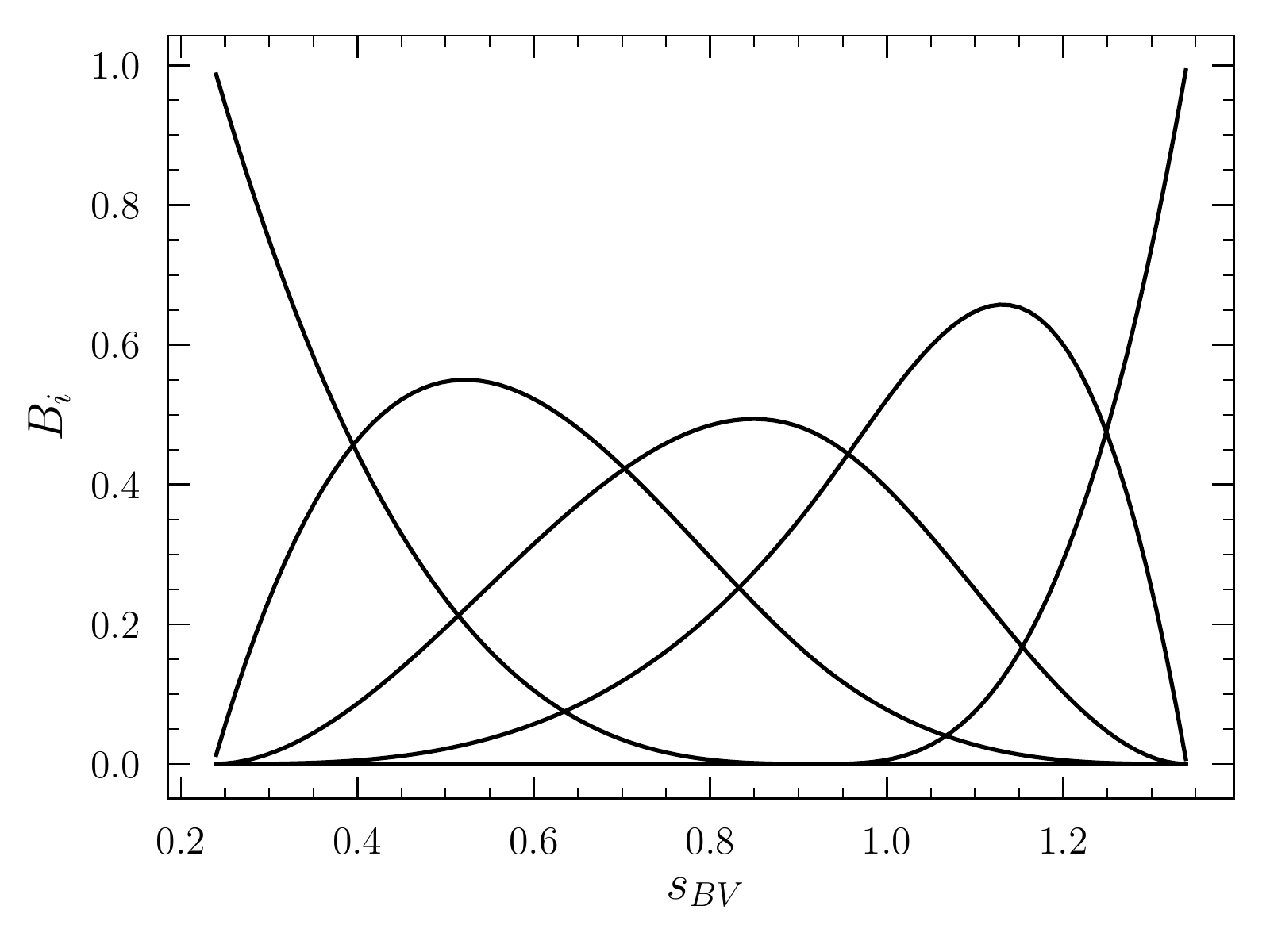}
\caption{The five basis splines used to model the intrinsic colors of the 
SNe~Ia. \label{fig:Bsplines}}
\end{figure}

We chose knot points at $s_{BV} = [0.23, 0.9, 1.34]$, corresponding to the two 
endpoints of the distribution and the median value. For a cubic spline, this
leads to 5 basis splines and therefore 5 degrees of freedom for each filter.
Figure \ref{fig:Bsplines} shows the functional form of these basis functions. Table
\ref{tab:splines_coef} lists the coefficients for several intrinsic colors of
SNe~Ia as well as the intrinsic scatter $\sigma$ in each color. One can construct
the spline coefficients for a color not found in Table \ref{tab:splines_coef} by
simply adding or subtracting the appropriate coefficients for two known colors.

\capstartfalse
\begin{deluxetable*}{lrrrrrr}
\tabletypesize{\scriptsize}
\tablecolumns{7}
\tablecaption{B-spline coefficients for intrinsic colors of SNe~Ia
   \label{tab:splines_coef}}
\tablehead{%
   \colhead{Color} & \colhead{$a_0$} & \colhead{$a_1$} & \colhead{$a_2$} &
   \colhead{$a_3$} & \colhead{$a_4$} & \colhead{$\sigma$} }
\startdata
$u-B$ & $ 1.21(0.12)$ & $ 0.77(0.14)$ & $ 0.16(0.13)$ & $ 0.41(0.09)$ & $ -0.08 (0.15)$ & $0.099$ \\
$B-V$ & $ 0.68(0.05)$ & $ 0.21(0.05)$ & $-0.25(0.06)$ & $ 0.06(0.04)$ & $ -0.18 (0.08)$ & $0.029$ \\
$g-r$ & $ 0.43(0.07)$ & $-0.08(0.06)$ & $-0.24(0.07)$ & $-0.19(0.05)$ & $ -0.40 (0.10)$ & $0.013$ \\
$V-r$ & $ 0.13(0.05)$ & $-0.12(0.05)$ & $-0.06(0.05)$ & $-0.17(0.03)$ & $ -0.21 (0.07)$ & $0.019$ \\
$V-i$ & $ 0.02(0.08)$ & $-0.51(0.07)$ & $-0.65(0.07)$ & $-0.90(0.05)$ & $ -0.76 (0.09)$ & $0.035$ \\
$r-i$ & $-0.11(0.06)$ & $-0.39(0.06)$ & $-0.59(0.06)$ & $-0.73(0.04)$ & $ -0.55 (0.07)$ & $0.035$ \\
$V-Y$ & $ 0.50(0.11)$ & $-0.60(0.10)$ & $-0.30(0.11)$ & $-1.20(0.08)$ & $ -0.63 (0.16)$ & $0.041$ \\
$Y-J$ & $ 0.01(0.11)$ & $ 0.37(0.12)$ & $-0.18(0.11)$ & $ 0.30(0.07)$ & $ -0.03 (0.13)$ & $0.078$ \\
$J-H$ & $ 0.03(0.10)$ & $-0.31(0.12)$ & $-0.01(0.11)$ & $-0.46(0.07)$ & $ -0.10 (0.12)$ & $0.073$ \\
$V-J$ & $ 0.51(0.13)$ & $-0.23(0.13)$ & $-0.48(0.13)$ & $-0.90(0.09)$ & $ -0.66 (0.16)$ & $0.070$ \\
$V-H$ & $ 0.55(0.12)$ & $-0.54(0.11)$ & $-0.49(0.12)$ & $-1.36(0.09)$ & $ -0.77 (0.17)$ & $0.029$ \\
\enddata
\end{deluxetable*}
\capstarttrue

\section{D. CSP-I Zero-points and their Uncertainties}
\label{sec:zero_points}
In order to measure $H_0$ using SNe~Ia, a sample of nearby objects
whose hosts have independent distances is required. Due to the scarcity of such
events, we must combine our CSP-I sample with others from the literature and so
a significant systematic uncertainty is the error in the absolute zero-points of
our photometric system. This was not presented in \citet{Krisciunas:2017}, so we
devote this section to their estimation. They apply equally well to both the CSP-I
and CSP-II filter sets.

\begin{figure}
   \plotone{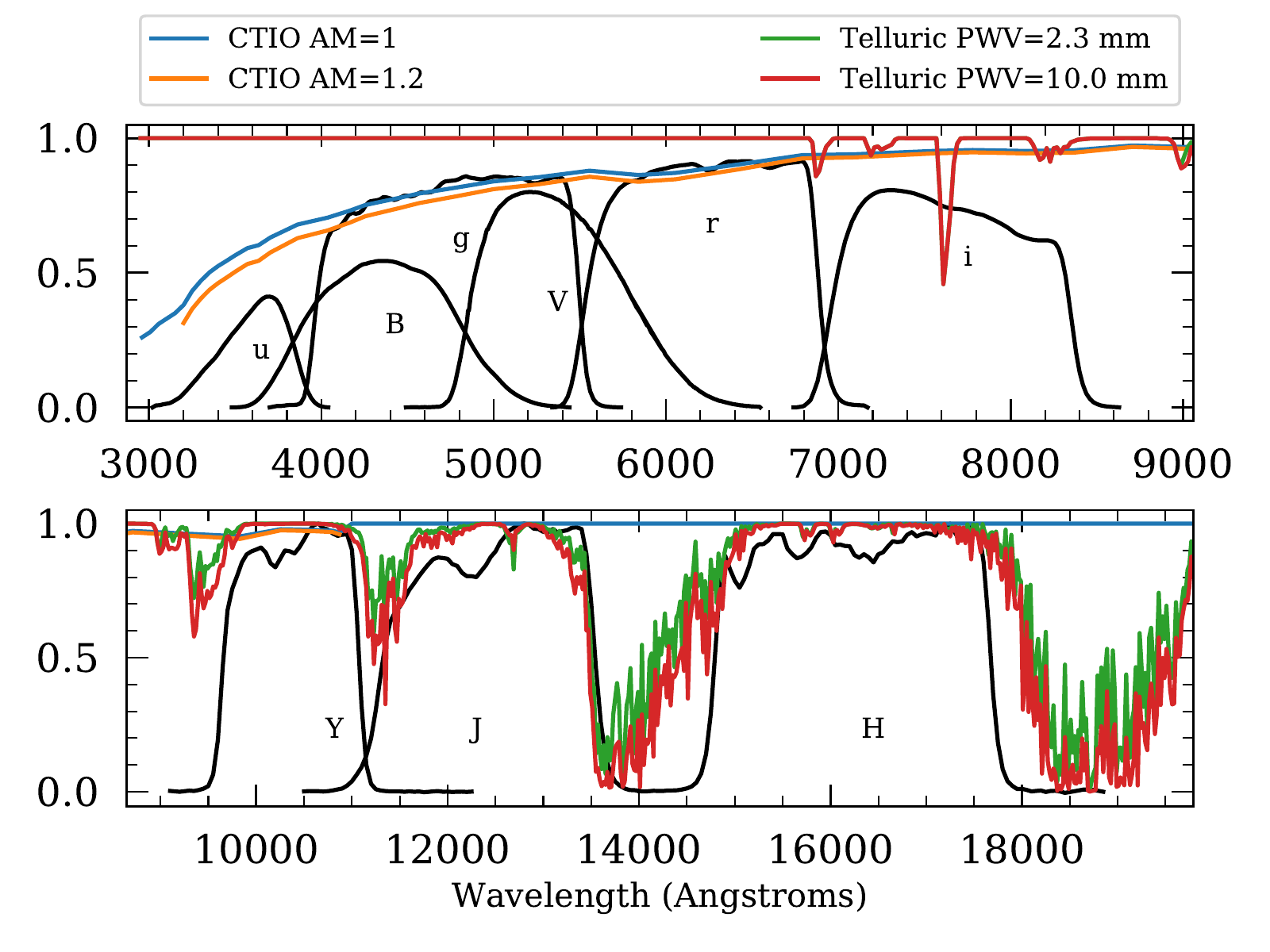}
   \caption{Scanned filter+telescope+CCD transmission lines (black curves) for
      optical (top panel) and NIR (bottom panel) bandpasses. The transmission
      due to atmospheric effects of aerosols (blue and orange lines) and
      telluric lines (red and green) are plotted for two different airmasses.
   \label{fig:filters}}
\end{figure}

The photometric zero-point for filter X allows us to convert from magnitudes to
photon fluxes measured by the telescope and is computed as:
\begin{equation}
zp_X = m_s + 2.5\log_{10}\left(\frac{1}{ch}\int f_{\lambda,s} S_{X} \lambda d\lambda 
             \right),
\label{eq:zpt}
\end{equation}
where $zp_{X}$ is the zero-point for filter $X$, $m_{s}$ is the assumed 
\emph{natural} magnitude
of a fundamental standard \citep{Krisciunas:2017}, 
$ch = 1.9864\times 10^{-8}$ erg $\cdot$ \AA,
$f_{\lambda,s}$ is the spectral energy distribution (SED) of the fundamental
standard in $\mathrm{erg\cdot s^{-1} \cdot cm^{-2} \cdot \AA^{-1}}$, $\lambda$
is the wavelength in \AA, and $S_{X}$ is the total 
transmission (telescope, instrument, and atmosphere) for filter $X$.

There are several sources of error in equation \ref{eq:zpt}, which we will now
estimate and combine in quadrature to get a final error in each $zp_X$. First,
there is the choice of fundamental standard. Typically the \citet{Bohlin:2004}
SED of Vega is used for
the set $B$, $V$, $Y$, $J$, $H$, while the \citet{Bohlin:2004a} SED of
$\mathrm{BD+17\degr 4708}$ is used for $u$, $g$ , $r$, and $i$. There are two other standards from
CALSPEC
\footnote{\href{http://www.stsci.edu/instruments/observatory/cdbs/calspec.html} {http://www.stsci.edu/instruments/observatory/cdbs/calspec.html}
} that have $B$- and $V$-band magnitudes on the \citet{Landolt:1992} system as 
well as $u$, $g$, $r$, $i$ magnitudes on the \citet{Smith:2002} system and so
can be used for our optical
filters:  P177D and P330E \citep{Bohlin:2015}. Indeed, 
$\mathrm{BD+17\degr 4708}$ is now known to be variable \citep{Bohlin:2015}.
Using these three different fundamental standards leads to shifts in the
zero-points listed as $\delta$(stand) in Table~\ref{tab:zpt_errors}. 
For the NIR, there is only
one other CALSPEC standard that has $J$ and $H$ photometry on the 
\citet{Persson:1998} system: $\xi^2$ Ceti (HD$15318$), which was one of the 
standards from \citet{Frogel:1978} used by \citet{Elias:1982} to calibrate 
their standards, and were in turn used to calibrate the
\citet{Persson:1998} standards. In the case of the $Y$ band, we use equation D1 from
\citet{Krisciunas:2017} to transform the $J$- and $H$-band magnitudes of
$\xi^2$ Ceti to a $Y$-band magnitude.

Second, in order to compute the natural system magnitudes $m_s$, we employ the 
color-terms from \citet{Krisciunas:2017} in reverse.  These color terms have
associated errors, leading to uncertainty in $m_s$, which are listed 
as $\delta (ct)$ in Table
\ref{tab:zpt_errors}.

Lastly, there remains the error in the shape of the transmission functions 
$S_{X}(\lambda)$. As detailed in \cite{Krisciunas:2017}, these are constructed
by multiplying the throughput of the telescope, instrument, and filters as
measured by \cite{Rheault:2010} with a model of the atmosphere:
\begin{equation}
S_X(\lambda) = R_{scan}(\lambda) A_{aero}(\lambda) 10^{-0.4 k_X (AM-1)} 
               A_{tel}(\lambda).
\label{eq;trans}
\end{equation}
Here $R_{scan}$ is the scanned throughput of the telescope, instrument and filter,
$A_{aero}$ is the extinction in the atmosphere due to aerosols and
ozone as measured at
CTIO for an airmass of 1.0 \citep{Stone:1983,Baldwin:1984}, 
$k_X$ is the extinction coefficient for
filter $X$, $AM$ is the typical airmass for the CSP standard observations, and
$A_{tel}$ is a model transmission function due to oxygen and water vapor. The 
most important systematic uncertainty in $R_{scan}$ is the wavelength 
calibration, which is estimated at $\pm 1$ \AA\ \citep{Rheault:2010}. Using 
synthetic photometry, we
found that the resulting error in the zero-points was negligible and we
therefore ignore any errors in $R_{scan}$ and consider only systematic
errors in the atmospheric transmission.

Figure \ref{fig:filters} shows the scans $R_{scan}$ and the atmospheric 
components for two different assumed values of airmass and precipitable
water vapor (PWV). It is obvious that errors in $A_{aero}$ are more
important for the optical filters, whereas errors in $A_{tel}$ is
more important in the NIR, including $i$ band. These errors are primarily 
due to the assumed
value of typical airmass $AM$ and PWV. As stated in \citet{Krisciunas:2017},
we assume $AM = 1.2$, which corresponds to the mode of the distribution
of $AM$ for our standard star observations. However, the distribution
has a long tail to high $AM$ and one could argue that the median
of $AM$ is more appropriate. The median for the CSP-I observations is
$AM = 1.33$ and we therefore take $d(AM) = 0.1$ as a possible systematic 
error in the assumed airmass. Using equation \ref{eq:zpt}, the errors in 
the zero-points due to errors $\delta(AM)$ and $\delta(K_S)$ are:
\begin{equation}
\delta(zp_Z) = \sqrt{k_X^2 \delta^2(AM) + AM^2 \delta^2(k_X)},
\end{equation}
and are listed in Table \ref{tab:zpt_errors}.

We now consider errors in $A_{aero}(\lambda)$ and $A_{tel}(\lambda)$.
Only errors that are correlated over significant portions of the filters will
be of any consequence. For $A_{aero}(\lambda)$, we have measurements 
of $k_X$ as part of our CSP-I photometry and since 
$k_X \sim -2.5\log_{10}A_{aero}(\lambda_X)$, we have sampled it at 9
wavelengths. To model a continuous function, we fit a Gaussian Process
to our observed $k_X$ using the CTIO curve as a mean function. This allows us
to generate a random sample of continuous extinction curves that are
consistent with our measurements and give a representative variation in
zero-point due to  changes in
the shape of the filter functions. These are listed in Table
\ref{tab:zpt_errors} as $\delta(A_{aero})$.

\capstartfalse
\begin{deluxetable*}{llllllll}
\tablewidth{0pc}
\tablecolumns{8}
\tablecaption{Photometric zero-point errors\label{tab:zpt_errors}}
\tablehead{
   \colhead{Filter} & \colhead{$\delta(\mathrm{stand})$} & 
   \colhead{$\delta(ct)$} & \colhead{$\delta(A_{aero})$} & 
   \colhead{$\delta(AM)$} & 
   \colhead{$\delta(k_X)$} & \colhead{$\delta A_{tel}$} &
   \colhead{$\delta(zp_X)$}  }
\startdata
$u$          & 0.004  & 0.016  & 0.073  & 0.051 & 0.012  & $\cdots$  & 0.091\\
$g$          & 0.005  & 0.003  & 0.013  & 0.019 & 0.005  & $\cdots$  & 0.024\\
$r$          & 0.001  & 0.002  & 0.012  & 0.010 & 0.004  & 0.001   & 0.016\\
$i$          & 0.005  & 0.002  & 0.021  & 0.006 & 0.004  & 0.002   & 0.022\\
$B$          & 0.005  & 0.000  & 0.018  & 0.024 & 0.005  & $\cdots$  & 0.031\\
$V$(LC-3014) & 0.003  & 0.004  & 0.011  & 0.014 & 0.004  & $\cdots$  & 0.019\\
$V$(LC-3009) & 0.006  & 0.004  & 0.011  & 0.014 & 0.004  & $\cdots$  & 0.020\\
$V$(LC-9844) & 0.003  & 0.004  & 0.011  & 0.014 & 0.004  & $\cdots$  & 0.019\\
$Y$          & 0.040  & $\cdots$ & $\cdots$ & 0.001 & $\cdots$ & 0.006   & 0.040\\
$J$          & 0.030  & $\cdots$ & $\cdots$ & 0.006 & $\cdots$ & 0.025   & 0.040\\
$H$          & 0.020  & $\cdots$ & $\cdots$ & 0.002 & $\cdots$ & 0.008   & 0.022\\
\enddata
\end{deluxetable*}
\capstarttrue

Finally, variations in $A_{tel}(\lambda)$ are very complicated and are 
primarily caused by changes in airmass and PWV. To model this, we use
ATRAN \citep{Lord:1992} models provided by Gemini Observatory for a variety
of $AM$ and PWV. For each, we build new filters and 
determine the effects on the zero-points. As above, we allow $AM$ to
vary by $\pm 0.1$. For the PWV, we utilize site testing data for the
Giant Magellan Telescope (GMTO) \citep{Thomas-Osip:2010}, 
which shows both nightly and seasonal
variations of PWV. We find a median of $\mathrm{PWV} = 4.3 $mm with seasonal
shifts of about 2 mm. We therefore assume the ``typical" PWV could be
in error by $\pm 2$mm. The effects of these variations in $AM$ and
PWV mostly affect the NIR filters and in particular the $J$ band, whose blue
and red edges extend well into the water bands (see Figure~\ref{fig:filters})
and whose shape is therefore influenced the most by PWV. The combined
effects are listed in Table \ref{tab:zpt_errors} as $\delta(A_{tel})$.

The final tallied error in the zero-points is listed in the final column
of Table \ref{tab:zpt_errors} and are used to model the systematic errors
when combining our photometry with other systems. All filters, aside from
$u$ are uncertain at the few percent level. The larger error in $J$ is mostly
due to the uncertainty in the effects of telluric absorption by the atmosphere
changing the shape of the filter and there is no clear way to reduce
these errors without measuring PWV, which was never done during the CSP-I
observations. The error in $Y$ has the potential to be greatly reduced by
improving the fundamental calibration. There are plans to build a small
($\sim 10$ inch aperture) robotic telescope at LCO with NIR capabilities,
which would allow us to observe brighter stars and tie the $Y$ band onto
the Vega system.

\section{E. Bayesian Hierarchical Models}
\label{sec:priors}
In this appendix we present some of the more technical aspects of our 
inference using Bayesian hierarchical models. A more general discussion
of the numerical method used (MCMC) can be found in our earlier work on
the intrinsic colors \citep{Burns:2014}. We restrict discussion here to
the model and priors used.

The models presented in sections \ref{sec:Intrinsic-Luminosities} and
\ref{sec:Hubble} are ultimately compared with observed 
fluxes $f_{\lambda,i}$ from the telescope. 
While the
models were presented in magnitude units to retain their familiarity,
all fitting is done in flux units. Conversion between the two is done
using the standard formula:
\begin{equation}
   f_{\lambda,i} = \exp\left[ 0.921\left(zp_\lambda - m_{\lambda,i}\right)\right]
\end{equation}
where $zp_{\lambda}$ is the photometric zero-point for the telescope and filter
in question. The precise values of $zp_{\lambda}$ are not important, as they
amount to a constant change of scale (akin to a change of units) and we 
choose values that produce fluxes as close to unity as possible, improving
numerical accuracy.

There is, however, a crucial difference between the probabilistic treatment of
the Cepheid and SN photometry. Typically, fluxes measured by a photon-counting
device such as a CCD are distributed as a Poisson distribution and for high
enough counts approach a normal distribution. The practice of converting to
magnitudes will result in a non-normal distribution with skew toward fainter
fluxes. This is the case for the SN~Ia data, where one can observe the host
galaxy after the SN has faded and subtract its flux entirely, leaving only
the SN flux, and a normal distribution is appropriate for 
describing it.

For the Cepheid data, the background host light
cannot be subtracted in such a way and its contribution must be estimated
in a probabilistic way by injecting multiple fake stars into the Cepheid
field, measuring their recovered flux, and comparing to the injected flux
\citep{Riess:2011}.
The resulting distribution of fluxes is dominated by fluctuations in the
background galaxy light, which are found to be normally distributed in
{\em magnitudes}\footnote{Adam Riess, private communication}, implying the 
fluxes are log-normally distributed. Log-normal distributions have a 
skew toward {\em brighter} fluxes and if the maximum likelihood for the
Cepheid photometry in 
magnitude units is $\hat{m}$, then the maximum likelihood in flux units will be
$\exp(\hat{m} + \sigma^2)$, where $\sigma$ is the observed standard deviation 
of the fake star photometry in magnitudes. The net result is a bias of order
$\sigma^2$. 
As part of their analysis, \citet{Riess:2016} removed this bias (and other
effects) from the published magnitudes, which amounts to multiplicative 
corrections in flux units and therefore biases the maximum likelihood in
flux to {\em fainter} values. Using a normal distribution in this case would
therefore lead to an underestimate of the flux and since the bias corrections
applied to the Cepheid data are not published, one must use a log-normal
distribution (or equivalently, fit in magnitude units).

The probability of the observed Cepheid fluxes $f_{ceph,i}$ given the model 
parameters $\theta$ is therefore
\begin{equation}
\label{eq:P_ceph_mobs}
P\left(f_{i,ceph}\vert \theta\right) = 
\mathrm{lognormal}\left(f_{i,ceph},f^T_{i,ceph}(\theta),\sigma^2_{i,ceph}\right)
\end{equation}
where $f^T_{i,ceph}(\theta)$ are the ``true'' Cepheid fluxes given by equations
\ref{eq:SN_Ceph} and \ref{eq:Cepheids}. The total variance $\sigma^2_{i,ceph}$
is modeled as the sum:
\begin{equation}
   \sigma^2_{ceph,i} =
\epsilon^2_i + \sigma^2_{ceph} + \gamma^2 \sigma^2_{OH}.
\end{equation}
where $\epsilon^2_i$ is the measurement error for $f_{i,ceph}$, 
$\sigma_{OH}$ is the error in $[O/H]$ which is treated as a 
free parameter, and 
$\sigma^2_{ceph}$ is any extra variance required to explain the
observed dispersion in the Leavitt law residuals.
It is assumed that 
the errors in $f_{ceph,i}$ are
uncorrelated with errors in $P_i$ or $[O/H]_i$ for the Cepheids
and that errors in period $P_i$ are negligible.
However, in order to allow the model to down-weight Cepheid variables based on 
their period \citep{Freedman:2001}, we add an extra error term
to $\sigma^2_{ceph}$:
\begin{equation}
   \sigma^2_{ceph} = \sigma^2_{ceph} + \exp\left(-(P-P_0)/P_{low})\right)
+ \exp\left(-(P_1-P)/P_{high}\right),
\end{equation}
where $(P_0,P_1)$ is the range of observed periods, and $P_{low}$ and 
$P_{high}$ control the lower and upper limits where the extra dispersion 
becomes important. The reason for doing this, instead of simply setting
hard limits, is that STAN needs to compute derivatives of the probability
with respect to the parameters and we therefore need an analytic (rather than
boolean) representation of the variance.

For the SNe~Ia, we assume the errors in $f_{\lambda,i}$
and $z_{hel,i}$ are uncorrelated. We use the measured covariance
between $f_{\lambda,i}$ and $s_{BV,i}$ from the SNooPy  fits
and assume normally-distributed errors in flux:
\begin{equation}
\label{eq:P_SN_mobs}
   P\left(\left[f_{\lambda,i},s_{BV,i}\right] \vert \theta\right) =
            N^2\left(\left[f^T_{\lambda,i}-f_{\lambda,i},s^T_{BV,i}-s_{BV,i}\right],
         \mathcal{C}_i\right),
\end{equation}
where $N^2$ is a 2D multivariate normal distribution centered at $[0,0]$
and with covariance matrix $\mathcal{C}_i$, 
$f^T_{\lambda,i}$ is the true value given by equations 
\ref{eq:Tripp_model} and \ref{eq:reddening_model}, and $s^T_{BV,i}$ are
the true color stretch values, which are nuisance parameters. 
As with the Cepheid data, we include a free parameter to model any
intrinsic variance $\sigma^2_{SN,\lambda}$ required to describe the
observed dispersion in the luminosity-decline-rate relation. We also
include a term that takes into account the added variance due to 
peculiar velocities of the host galaxies that arises when using 
redshift to compute distances. These terms are added
to the $(0,0)$ component of $\mathcal{C}_i$:
\begin{equation}
   \mathcal{C}_i(0,0) = \epsilon^2_{\lambda,i} + \sigma^2_{SN,\lambda} +
                        \left(\frac{v_{pec}}{cz_{cmb}}\right)^2,
\end{equation}
where $\epsilon_{\lambda,i}$ is the measurement error in $f_{\lambda,i}$
and $v_{pec}$ is the typical peculiar velocity for a SN~Ia host, which
we treat as a free parameter.

The probability for the entire dataset $D$ is then the product:
\begin{equation}
   \label{eq:P_data}
   P\left(D\vert \theta\right) = \prod_i P(f_{ceph,i}\vert D) 
                              \prod_j P(f_{\lambda,j}\vert D).
\end{equation}
Using Bayes' theorem, the posterior of the parameters of interest is
given by
\begin{equation}
   P(\theta\vert D) \propto P\left(D\vert \theta \right) P(\theta),
\end{equation}
where $P\left(D\vert \theta \right)$ is given by equation
\ref{eq:P_data} and $P(\theta)$ are the priors on our variables. 
Experimentation has 
shown that uniform priors are appropriate for all our nuisance parameters
save those that involve intrinsic variances and the Cepheid period range
($P_{low}$ and $P_{high}$), for which we impose strictly positive values
\footnote{Note that uniform priors placed on parameters expressed in 
magnitudes implies Jeffreys priors in flux units.}.
For the extinction parameters of the SNe~Ia, $E(B-V)_i$ and $R_{V,i}$, 
we use priors based on the results of the color-based analysis detailed
in section \ref{sec:color_reddening}. A summary of all the parameters and their
typical values and errors is given in Table \ref{tab:parameters}.

\clearpage
\bibliographystyle{aasjournal}
\bibliography{AnalysisII.bib}

\end{document}